\documentclass[runningheads]{llncs}
\usepackage[T1]{fontenc}
\usepackage{graphicx}
\usepackage{url}
\usepackage[dvipsnames]{xcolor}
\usepackage{comment}

\usepackage{tikz}
\usetikzlibrary{arrows.meta, positioning}
\usetikzlibrary{arrows.meta, shapes.geometric, positioning}
\tikzstyle{circleproc} = [circle, draw=black, fill=blue!10, minimum size=2.8cm, text centered]
\tikzstyle{arrow} = [thick, ->, >=stealth]

\usepackage{tabularx}
\usepackage{caption}
\usepackage{geometry}
\usepackage{mdframed}
\usepackage{framed}

\begin{document}

% \title{Challenges Faced by CS Educators: A Thematic Review of Educator Challenges\thanks{Supported by organization x.}}
% A Thematic Review of CS Educator Challenges
% \title{Challenges Faced by CS Educators: \\A Systematic Mapping Study of Educator Challenges}
\title{CS Educator challenges and their solutions: \\ A systematic mapping study}

% \author{
% First Last\orcidID{xxxx-Xxxx-xxxx-xxxx} \and
% First Last\orcidID{xxxx-Xxxx-xxxx-xxxx} \and
% First Last\orcidID{xxxx-Xxxx-xxxx-xxxx}
% }

% % \author{
% % Anjali Chouhan\orcidID{0009-0002-7994-1374} \and
% % Sruti Srinivasa Ragavan\orcidID{0000-0001-6197-2194} \and
% % Amey Karkare\orcidID{0000-0002-3664-6490}
% % }

% \authorrunning{anonymous et al.}
% \institute{Anonymous affiliation \\
% \email{\{email1, email2, email3\}@abc.de.fg}
% }
% % \authorrunning{A. Chouhan et al.}
% % \institute{Indian Institute of Technology Kanpur, Uttar Pradesh, India \\
% % \email{\{anjalic24, srutis, karkare\}@cse.iitk.ac.in}
\author{
Anjali Chouhan\orcidID{0009-0002-7994-1374} \and
Sruti Srinivasa Ragavan\orcidID{0000-0001-6197-2194} \and
Amey Karkare\orcidID{0000-0002-3664-6490}
}

\authorrunning{A. Chouhan et al.}

\institute{Indian Institute of Technology Kanpur, Uttar Pradesh, India \\
\email{\{anjalic24, srutis, karkare\}@cse.iitk.ac.in}
}
\newcommand{\todo}[1] {\textbf{\textcolor{red}{#1}}}

\maketitle

\begin{abstract}
Computer Science (CS) education is expanding rapidly, but educators continue to face persistent challenges in teaching and learning environments.Despite growing interest, limited systematic work exists to categorize and synthesize the specific challenges faced by CS educators and the remedies adopted in response.This is problematic because it remains unclear which areas have been thoroughly addressed and which still lack sufficient scholarly attention. 

In this study, we conducted a structured literature review of peer-reviewed research papers published over the last five years, focusing on challenges and remedies across ten categorized themes, including pedagogical, emotional, technological, and institutional dimensions.Our analysis revealed recurring issues in areas such as assessment practices, teacher training, classroom management, and emotional well-being, along with various strategies—such as professional development programs and policy interventions—adopted to mitigate them while also revealing several areas that have received insufficient attention.This review offers a consolidated understanding of the CS education landscape, providing valuable insights for researchers, curriculum designers, and policymakers aiming to improve teaching effectiveness and educator support.


% \keywords{Computer Science Education, CS Educators, Teaching Challenges, Pedagogical Barriers, Teaching Assistants (TAs), Literature Review, Educational Remedies, Professional Development, Institutional Challenges, Emotional Well-being, Classroom Management}

\end{abstract}

\section{Introduction}
% What is the context? What is the problem? What have others done? What is missing in that? How are we addressing the missing part? How are we doing that (briefly)? 

As Computer Science education (CSEd) continues to expand globally, instructors, teaching assistants (TAs), and faculty members play a critical role in shaping effective and inclusive learning environments. However, these educators often encounter a diverse set of challenges—pedagogical~\cite{Thigpen2024}, emotional~\cite{Farghally2024}, technological~\cite{Vivian2020} , and institutional~\cite{Valstar2020}—that can significantly affect teaching effectiveness and student outcomes.

Indeed, studies have explored specific challenges CS educators face in various contexts. For example, a multi-institutional study in India investigated the programming abilities of Computer Science (CS) instructors, particularly those from mid-tier institutions, revealing a sharp
fall in their ability to reason about code abstractly, modify given code, and write code~\cite{Kumar2022}, another multi-institutional
and multi-national study explored the challenges faced by Teaching Assistants (TAs) in computer science
education across Europe~\cite{Riese2021}. Others (e.g., \textit{UserFlow}\textit{Solvelets}~\cite{Singh2020,Kumar2022}, Plagiarism detection tools~\cite{Forden2023,Shrestha2022}, Code quality tools~~\cite{Izu2025}) have also proposed and evaluated interventions to address those challenges. However, there remains a lack of comprehensive synthesis  of these two bodies of work (e.g., via systematic literature review, systematic mapping studies) that captures the full range of difficulties faced by educators and the ways in which literature addresses them. This gap makes it difficult to: 1) gain a holistic, multi-dimensional understanding of various categories of educators' problems, 2) provide solutions that integrate various successful approaches that carefully consider potential tradeoffs, 3) develop foundational theories and 4) identify prominent areas where efforts have been made, and -more importantly- areas where none exists. 

To bridge this gap, we conducted a structured mapping study of peer-reviewed publications in the CS Education (CSEd) domain on educator challenges from the last five years. A systematic mapping study is a metastudy approach that systematically synthesizes literature in an area, with the specific aim of synthesizing focus areas, summarizing contributions and identifying gaps~\cite{Budgen2008Mapping}. Our investigation was guided by the following research questions: 

\begin{itemize}
\item[RQ1:] What are the key challenges that CS educators face, as reported in the CSEd literature from the last five years?
\item [RQ2:] What remedies or interventions have been proposed or implemented in response to these challenges, as reported in the literature?
\item [RQ3:] Where are the key gaps where for CSEd research community can better assist CS educators?
\end{itemize}

Synthesizing a total of 106 relevant studies, we were able to answer the above questions. Our key findings include ten broad categories of challenges CS educators face--of which, only 5 have received attention in the literature in the form of solution approaches . Our results offers insights to guide future research, policy development, and support mechanisms for CS educators.

\section{Background and Related Work}

\subsubsection{Who Are CS Educators?}

To investigate challenges CS educators face, we have to operationalize the term CS educator. In this discourse, we define CS educators as any person who teaches, mentors, or supports student learning in various learning environments and contexts. These include:
\begin{itemize}
    \item \textit{Instructors and Faculty}, who design and/or teach diverse CS courses—ranging from CS1~\cite{Vahid2024,Vahid2024b}, capstone projects~\cite{Hooshangi2025}, and ethics-focused courses~\cite{Hu2023,Parthasarathy2024} to K–12 education~\cite{Grover2024,Jetzinger2024,DiPaola2023}. They have different roles and challenges based on their contexts. Some also influence computing standards~\cite{Thigpen2024,McGill2024}, lead reforms~\cite{Musaeus2024}, and mentor future educators~\cite{Farghally2024}--playing larger roles beyond simply teaching.
 
    \item \textit{TAs} support instruction through grading, feedback, and conceptual scaffolding~\cite{Zaman2023,Akhmetov2024,Haglund2024}, often enhanced via tools like TA-Bot~\cite{Forden2023,Denny2024}. Addressing issues of TAs, especially in training,  feedback literacy and scaling to large classes, are vital for CS education research~\cite{Cheng2024}.
   
    \item \textit{Tutors} are individuals that provide targeted academic support to learners, often supplementing classroom instruction. They assist with reinforcing concepts, offering feedback, and guiding students through assignments or problem-solving activities. Traditionally, tutors have been human—such as peer mentors or academic assistants—but increasingly, AI-powered systems are fulfilling similar roles. Tutors offer personalized support, as do their AI counterparts~\cite{Bassner2024,Malik2024}. In general, just-in-time personalized training has been shown to enhance the efficacy and relevance of tutoring interventions~\cite{Cheng2024}.
    
     % \textbf{Tutors:} Tutors, whether human or AI-driven, offer one-on-one or small-group support for learners. Systems like \textit{Iris}, an AI-based tutor, provide real-time, context-aware support that fosters independent problem-solving~\cite{Bassner2024}. Similarly, tools like \textit{TeachNow} enable volunteer teachers to deliver spontaneous 1:1 assistance in massive online courses~\cite{Malik2024}. Tutors are also supported by just-in-time interventions that improve their instructional performance~\cite{Cheng2024}.
    
    % \textbf{K–12 and Pre-Service Teachers:} These educators face unique hurdles in PD~\cite{Shanley2023,Macann2023,DiPaola2023,Jetzinger2024}, curriculum design~\cite{Grover2024,Greifenstein2023}, and ethics integration~\cite{Hu2023}, shaping early CS education.
    \item \textit{K-12 and Pre-Service Teachers:} A significant portion of computing educators operate in primary and secondary school settings.Their perspectives shape the landscape of early computing education.
\end{itemize}

% These educators face unique challenges related to professional development~\cite{Shanley2023,Macann2023,DiPaola2023,Jetzinger2024}, curriculum development~\cite{Grover2024,Greifenstein2023}, and ethics integration~\cite{Hu2023}. 

% This collective includes educators operating across modalities, content areas, and learner levels.

% A substantial and expanding body of research in Computer Science Education (CS Ed) addresses diverse dimensions of teaching and learning for both educators and students. Despite this progress, persistent gaps remain—particularly in the areas of sustained professional development and comprehensive support systems for CS educators.

\subsubsection{Empirical studies of CS educators}
Empirical studies extensively investigate various aspects of computer science education, often focusing on teachers' practices, experiences, and professional development needs across different educational levels and geographical contexts. For example, a multi-institutional study in India investigated the programming abilities of Computer Science (CS) instructors, particularly those from mid-tier institutions, revealing a sharp fall in their ability to reason about code abstractly, modify given code, and write code~\cite{Kumar2021}. This study focused primarily on CS instructors at the university level. Another study engaged 34 educators from across the United States in an AI Educator Make-a-Thon to empower middle school teachers as AI curriculum designers and leaders, demonstrating a statistically significant increase in their sense of belonging and self-confidence in curriculum design~\cite{DiPaola2023}. Research from China surveyed 164 primary and secondary school teachers on their Scratch programming practices, finding that teachers' self-efficacy significantly correlated with their use of integrated direct instruction and classroom management strategies~\cite{Jia2022}. Furthermore, a multi-institutional and multi-national study explored the challenges faced by Teaching Assistants (TAs) in computer science education across Europe, analyzing 180 reflective essays to identify five main challenges including becoming a professional TA and student-focused issues~\cite{Riese2021}. At the university level, a mixed-methods study provided global insights into capstone course instructors' perspectives, noting their high intrinsic motivation despite perceiving these courses as demanding significantly more time and effort than regular courses~\cite{Hooshangi2025}. These studies collectively highlight the diverse research approaches employed to understand and improve educators' roles in CS education from K-12 to university levels.

\subsubsection{Tools and interventions for supporting educators}
Significant efforts are dedicated to supporting computer science educators through various programs, tools, and professional development initiatives designed to enhance their skills, confidence, and teaching effectiveness. For instance, the TeachNow system enables volunteer teachers to provide spontaneous, real-time, one-on-one help in massive online courses (MOOCs), effectively leveraging fragmented volunteer time and significantly increasing student course retention rates~\cite{Malik2024}; more such are described later in this paper. In the context of India, a study identified that CS faculty need specific training on accessibility concepts and disabilities sensitization, along with exposure to existing CS education research and pedagogies. They also explicitly sought hands-on resources~\cite{Parthasarathy2024}. Practical strategies for training graduate CS Teaching Assistants (TAs) to provide effective feedback have been developed, addressing the specific needs of international TAs in low-resource and high-diversity university contexts~\cite{Zaman2023}. Comprehensive professional development (PD) programs have been shown to be crucial for successful adoption of complex pedagogies like Process Oriented Guided Inquiry Learning, extending beyond one-time workshops to include sustained mentoring and peer meetings~\cite{Kussmaul2022}. Furthermore, scalable PD offers have been designed for K-12 CS teachers in various countries to prepare them for Artificial Intelligence (AI) introduction into mandatory curricula~\cite{Jetzinger2024}. These support mechanisms aim to address diverse challenges faced by educators and ensure high-quality CS instruction across all levels.

\subsubsection{Meta studies and literature reviews of CS educator research}
The field of computer science education increasingly relies on literature reviews and meta-studies to synthesize existing knowledge and identify broader trends concerning teachers and their professional development. A comprehensive review of international models of Computer Science teacher education explored how CS teachers are prepared across various countries and states, including the US, Germany, Ireland, the Netherlands, New Zealand, and Spain, highlighting the significant global expansion of CS education and diverse approaches to teacher preparation, from short workshops to full degree programs~\cite{Yadav2022}. In the African context, the first comprehensive literature review of computing education research revealed a substantial body of work and identified the critical need for contextualization in learning materials for CS1 and other areas~\cite{Hamouda2025}. A longitudinal analysis specifically focused on seven years of K-12 computing education research in the United States, analyzing over 500 articles to identify major trends in curriculum, concepts taught, and student demographics, and noting concerns about underreporting of demographic data~\cite{Upadhyaya2020}. Furthermore, a global survey of nearly 100 introductory programming (CS1) instructors across 18 countries provided the first broadly-scoped comparison of global trends in CS1, examining reasons for language and environment choices and the prevalence of online delivery, offering a ``snapshot'' before the widespread influence of Generative AI tools~\cite{Mason2024}. Another extensive collection compiled diverse experiences of numerous academics regarding the use of automated assessment (AA) in tertiary computing courses, highlighting motivations like reducing workload, ensuring consistency, and supporting academic misconduct identification from the educators' perspectives~\cite{Luxton-Reilly2023}. These meta-studies provide crucial overviews, helping to inform evidence-based recommendations and guide future research directions in CS teacher education.
\subsubsection{Literature gaps and present work}
However, a major gap is the lack of a review of educator needs and challenges. Educators--instructors, TAs, tutors--at various levels are crucial in CS education, but face various challenges. The current focus on AI education and literacy has brought these challenges to the forefront, with challenges around curriculum development, content adaptation, pedagogical adoption and contextualization, and simply, teacher upskilling to learn and teach these new concepts. This calls for a review of various teacher challenges broadly, what has been done so far in this space, and where opportunities lie. 

Moreover, from our review of 1140 peer-reviewed publications in CS education, only 108 focused directly on the challenges faced by educators. This disparity further underscores the need for a more structured and comprehensive synthesis of how existing work addresses (or fails to address) real-world educator needs. Our study aims to close this gap by systematically mapping the landscape and identifying actionable areas for future research and intervention in supporting CS educators.

\section{Systematic Mapping Methodology}
\subsection{Literature Search}

To identify relevant research on challenges and training in computer science (CS) education, we conducted a systematic literature search using the ACM Digital Library. Our goal was to collect recent studies focused on CS educators in various categories and across various levels of education published in the recent years in top computing education venues. Therefore, we employed a structured query targeting keywords related to programming instruction and educator development, collaboratively crafted by all three authors. We began with the following query:

\begin{quote}
[[Abstract: programming] OR [Abstract: coding] OR [Abstract: cs1] OR
[Abstract: cs2]]
AND [[Abstract: teacher] OR [Abstract: educator] OR [Abstract: tutor]
OR [Abstract: teaching assistant]]
AND [[Abstract: training] OR [Abstract: pedagogy]]
AND [[Publication Title: SIGCSE] OR [Publication Title: TOCE] OR
[Publication Title: CompEd] OR [Publication Title: ICER]
OR [Publication Title: ITiCSE] OR [Publication Title: ACE]]
AND [E-Publication Date: Past 5 years]
AND [CCS 2012: Computing Education]
AND [CCS 2012: Computing Education Programs]
\end{quote}

This query, executed on February 14, retrieved 565 papers. On cross-validating the query and its retrieved papers for completion (by checking for inclusion of known and random papers found through Google Scholar Search), we found gaps and therefore expanded our search query to incorporate additional terms related to educator roles and challenges, and by broadening the scope of CCS classifications, resulting in the following query:

\begin{quote}
[[Abstract: programming] OR [Abstract: coding] OR [Abstract: cs1] OR
[Abstract: cs2] OR [Abstract: pedagogy]]
AND [[Abstract: teacher] OR [Abstract: educator] OR [Abstract: tutor]
OR [Abstract: teaching assistant]]
AND [[Abstract: training] OR [Abstract: pedagogy] OR 
[Abstract: challenges] OR [Abstract: struggles] 
OR [Abstract: problems] OR [Abstract: educating] OR 
[Abstract: learning]]
AND [[Publication Title: TOCE] OR [Publication Title: ITiCSE] OR 
[Publication Title: SIGCSE] OR [Publication Title: CompEd] 
OR [Publication Title: ICER] OR [Publication Title: ACE]]
AND [E-Publication Date: Past 5 years]
AND [CCS 2012: Computing Education Programs OR Computing Education]
\end{quote}

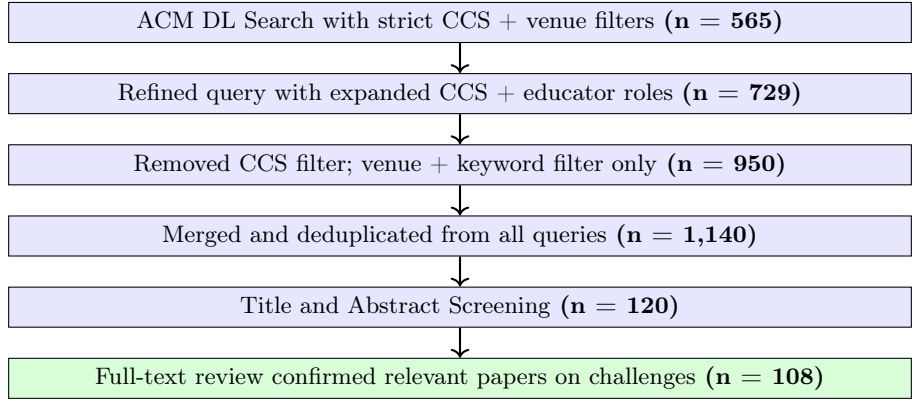
\begin{figure}[ht]
\centering
\begin{tikzpicture}[node distance=0.4cm, every node/.style={align=center}]

% Nodes
\node (start) [rectangle, draw, minimum width=12cm, minimum height=0.5cm, fill=blue!10] {ACM DL Search with strict CCS + venue filters~\textbf{(n = 565)}};

\node (step2) [rectangle, draw, minimum width=12cm, minimum height=0.5cm, below=of start, fill=blue!10] {Refined query with expanded CCS + educator roles~\textbf{(n = 729)}};

\node (step3) [rectangle, draw, minimum width=12cm, minimum height=0.5cm, below=of step2, fill=blue!10] {Removed CCS filter; venue + keyword filter only~\textbf{(n = 950)}};

\node (step4) [rectangle, draw, minimum width=12cm, minimum height=0.5cm, below=of step3, fill=blue!10] {Merged and deduplicated from all queries~\textbf{(n = 1,140)}};

\node (step5) [rectangle, draw, minimum width=12cm, minimum height=0.5cm, below=of step4, fill=blue!10] {Title and Abstract Screening~\textbf{(n = 120)}};

\node (final) [rectangle, draw, minimum width=12cm, minimum height=0.5cm, below=of step5, fill=green!15] {Full-text review confirmed relevant papers on challenges~\textbf{(n = 108)}};

% Arrows
\draw[->, thick] (start) -- (step2);
\draw[->, thick] (step2) -- (step3);
\draw[->, thick] (step3) -- (step4);
\draw[->, thick] (step4) -- (step5);
\draw[->, thick] (step5) -- (final);

\end{tikzpicture}
\caption{Paper selection and screening process.}
\label{fig:papers_filtering}
\end{figure}

The above query, retrieved 729 papers, but still failed some cross validation checks against known papers. Specifically, we found mismatch between CCS classification as indexed in the ACM Digital Library and what is mentioned in the paper. For example~\cite{Kumar2021} had been indexed under CCS classification—\textit{Professional Topics → Computing Profession}—instead of the expected \textit{Computing Education} category as the paper's PDF claims. This highlights a potential limitation of relying on CCS classifications or paper keywords for literature retrieval, as relevant work may occasionally be categorized under adjacent but non-obvious topics. Therefore, we removed removed the CCS filter entirely and conducted a broader search based on abstract keywords and publication venues alone; this resulted in 950 more papers. After merging the results from the multiple iterations and removing duplicates, we obtained a final corpus of 1,140 papers (Figure~\ref{fig:papers_filtering}).

\subsection{Paper Selection and Categorization}
% Figure 1: A dataflow diagram showing the number of papers
%  in our dataset at each step in the analysis.
\begin{itemize}
\item \textbf{Inclusion Criteria}: We included papers focusing on, or motivated by, teacher, TA, or tutors in CS education across all levels, in both formal and informal educational contexts.

\item \textbf{Exclusion Criteria}: We excluded papers that did not explicitly focus on educators (e.g., improving student experiences). We excluded papers focused on non-CS educators. (e.g., math education).

\item \textbf{Initial screening}: We began by reading the title and abstract of the 1140 papers in our corpus. Applying the above inclusion and exclusion criteria, we screened in 120 papers. 

\item \textbf{Full review}: We then conducted a full-text review of the 120 selected papers to evaluate its relevance. We found only 108 were in fact relevant to our theme of educator challenges and solutions. In this phase, we also went about extracting metadata (e.g., list of challenges, country of study) for further analysis.

\item \textbf{Reliability}: To ensure reliability in our process, a random sample of inclusion and exclusion of papers at each stage was reviewed by another researcher, namely the second author.
\end{itemize}

% \caption{Funnel diagram of paper selection and screening process used in the systematic mapping study.}
% \label{fig:funnel_process}
% \end{figure}
% \begin{figure}
%     \centering
%     \includegraphics[width=1.0\linewidth]{Images/funnelpython.png}
%     \caption{Funnel diagram of paper selection and screening process used in the systematic mapping study.}
%     \label{fig:funnelpython}
% \end{figure}

\subsection{Analysis}
We then categorized these papers into themes for further analysis. The first author categorized the papers into themes and based on metadata (target educator, nature of contribution, country, a list of challenges teachers faced, etc.) and the second author randomly reviewed the categorization.

\section{Findings}
We organize our findings around the three research questions listed earlier, with a prelude of the broader research trends in which to anchor and interpret them. 

\subsection{Broad trends in CS educator research}
\begin{figure}[ht]
    \centering
    \begin{minipage}{0.38\textwidth}
        \centering
        \includegraphics[width=\textwidth]{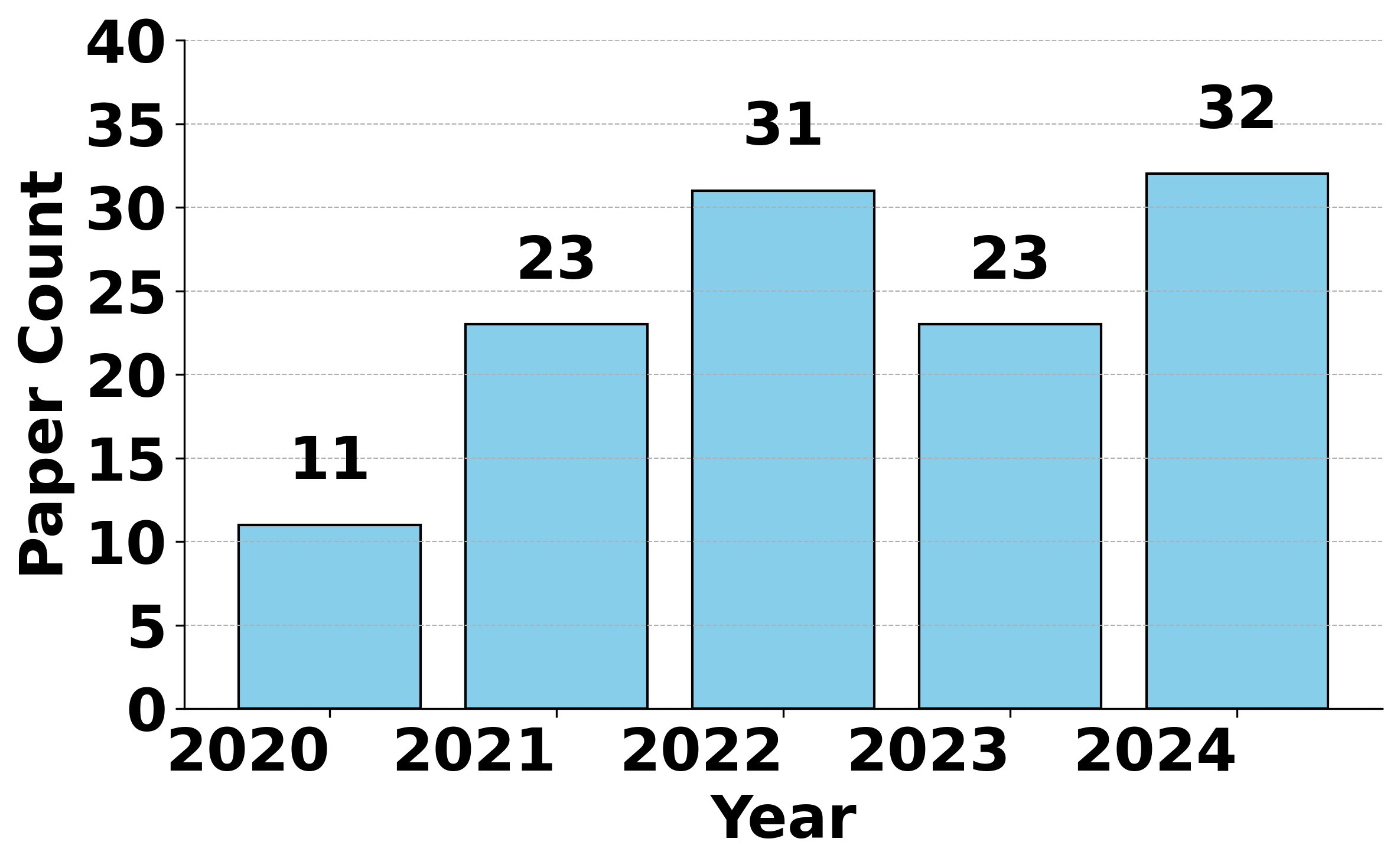}
        \caption{Year-wise Publication Count (2020--2024)}
        \label{fig:yearwise_publications}
    \end{minipage}\hfill
    \begin{minipage}{0.58\textwidth}
        \centering
        \includegraphics[width=\textwidth]{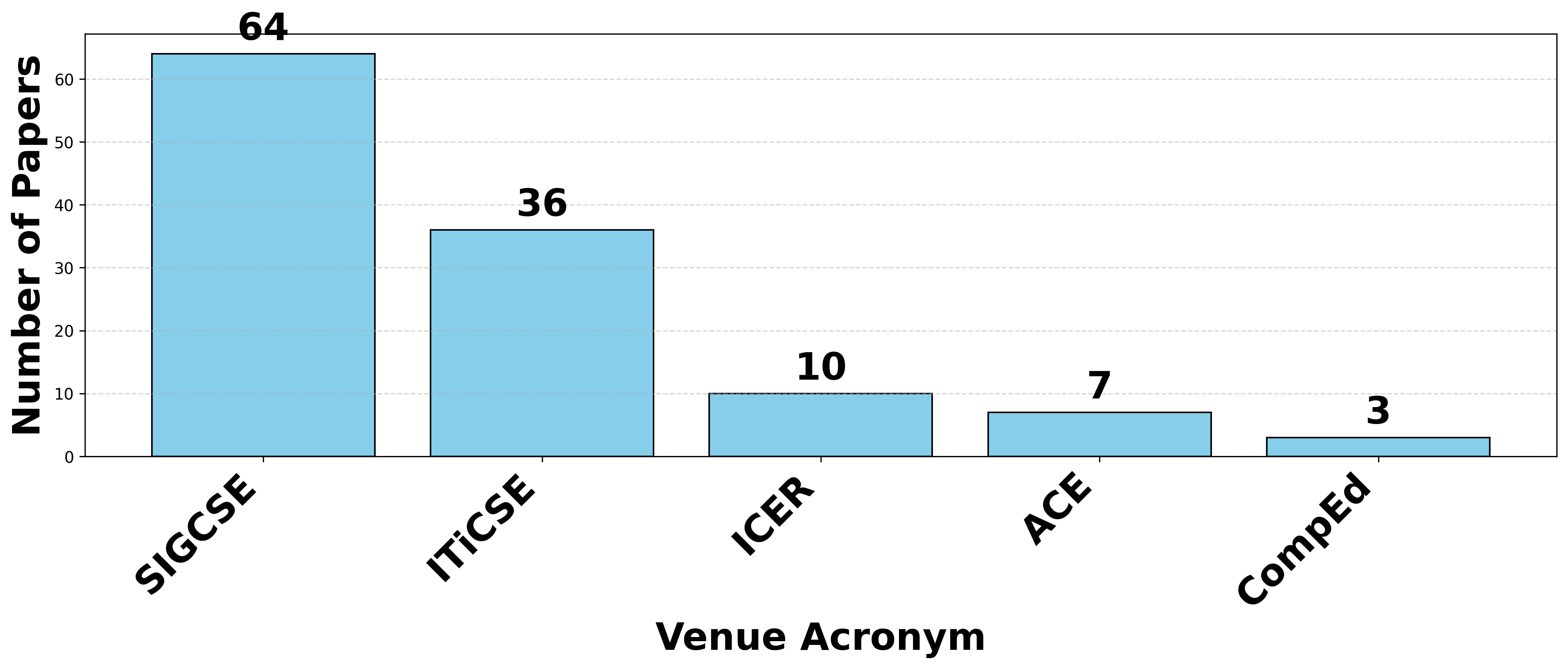}
        \caption{Distribution by Publication Venue}
        \label{fig:venue}
    \end{minipage}
\end{figure}

First and foremost, as Figure~\ref{fig:yearwise_publications} illustrates, the number of studies focusing on educators has steadily increased between 2020 and 2024. This upward trend underscores growing scholarly and institutional recognition of the challenges faced by CS educators in recent years.

% \begin{figure}[ht]
%     \centering
%     \includegraphics[width=0.7\textwidth]{Images/year2.2.png}
%     \caption{Year-wise Publication Count (2020--2024)}
%     \label{fig:yearwise_publications}
% \end{figure}

These papers mostly come from CS education conferences, most prominently SIGCSE Technical Symposium, ITiCSE (Figure~\ref{fig:venue}) and the newer CompEd. General education research conferences such as ICER and ACE also feature such teacher-focused works. 

% \begin{figure}[ht]
%     \centering
%     \includegraphics[width=0.7\textwidth]{Images/venue2.2.png}
%     \caption{Distribution by Publication Venue}
%     \label{fig:venue}
% \end{figure}

% \begin{figure}[h!]
%     \centering
%     \includegraphics[width=0.7\textwidth]{Images/nature2.1.png}
%     \caption{Distribution of Nature of Study}
%     \label{fig:nature}
% \end{figure}

\begin{figure}[htbp]
    \centering
    \begin{minipage}[b]{0.38\textwidth}
        \centering
        \includegraphics[width=\linewidth]{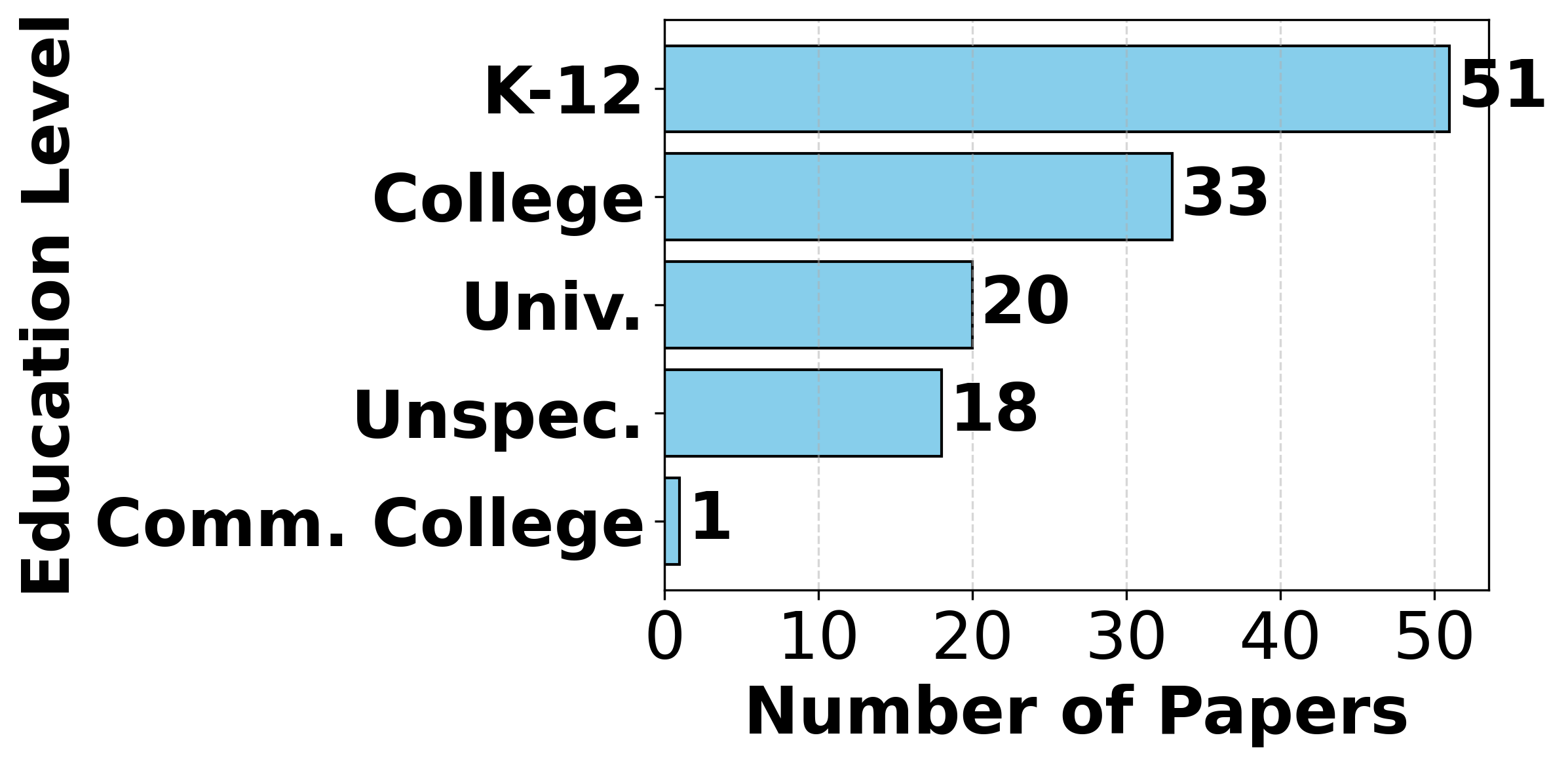}
        \caption{Distribution by Educational Level}
        \label{fig:edulevel}
    \end{minipage}
    \hfill
    \begin{minipage}[b]{0.58\textwidth}
        \centering
    \includegraphics[width=0.7\textwidth]{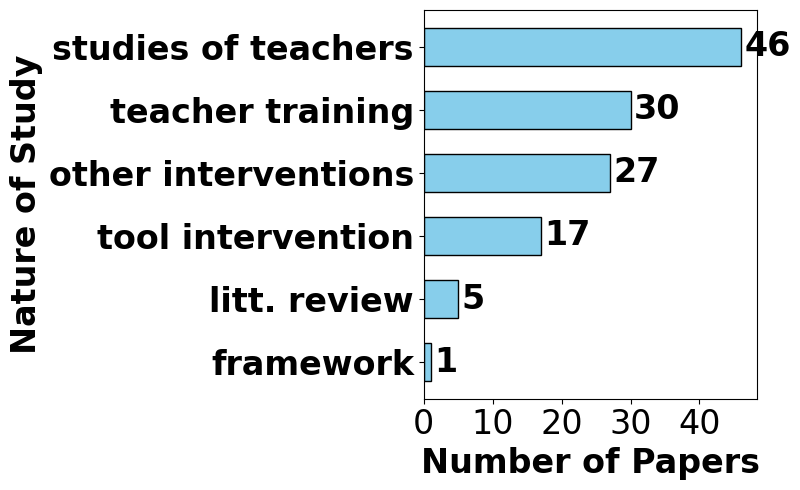}
    \caption{Distribution of Nature of Study}
    \label{fig:nature}
    \end{minipage}
\end{figure}

Coming to the nature of work, (Figure~\ref{fig:nature}) highlights the community's broad focus on empirical studies of teachers, presumably to understand their needs and challenges, and various interventions to address their problems; these two categories account for nearly a third of the papers each. Teacher training as a form of intervention has also received heavy focus. However, the area still remain less mature and efforts towards theorization or development of frameworks are few.
Thus, the first takeaway of our study is: 

\begin{framed}
    \noindent \textbf{Takeaway \#1:} As a community, there is opportunity for CSEd researchers to move beyond empirical work in silos to develop coherent theories and frameworks surrounding various aspects of educators.
\end{framed}

\begin{figure}[htbp]
    \centering
    \begin{minipage}[b]{0.48\textwidth}
        \centering
        \includegraphics[width=\linewidth]{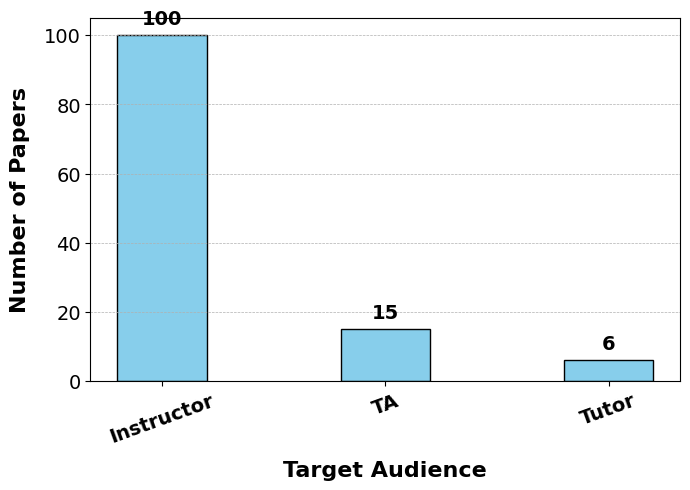}
        \caption{Target educator population}
        \label{fig:audience}
    \end{minipage}
    \hfill
    \begin{minipage}[b]{0.48\textwidth}
        \centering
        \includegraphics[width=\linewidth]{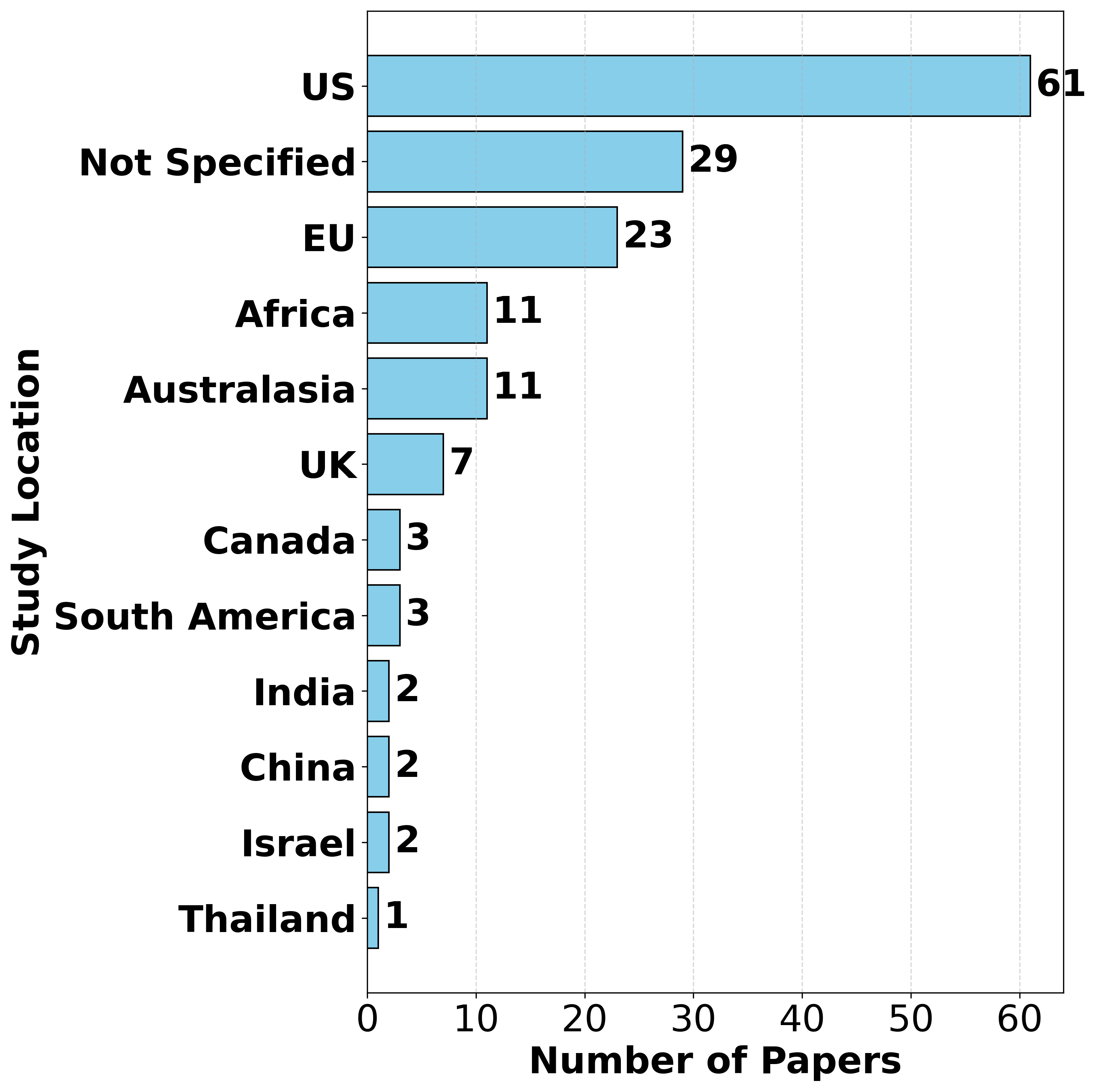}
        \caption{Country-wise Distribution of Studies}
        \label{fig:country}
    \end{minipage}
\end{figure}

There is also a general lack of diversity in terms of who the research mainly aims to serve. Geographically,(Figure~\ref{fig:country}) most papers are US centric with only about 20\% of papers considering educational contexts outside of Europe, UK, North Americas and Australasia. Nearly 30\% of papers also do not report the demographics of their study explicitly, calling for rigor in reporting and reviews. An overwhelming majority of research has also focused mostly on K-12 and college/university teachers (Figure~\ref{fig:audience}), with alternative education settings or community colleges receiving almost no attention (Figure~\ref{fig:edulevel}). Focus has also been on empowering instructors, overlooking TAs and tutors that are essential players in the educational ecosystem of a learner (Figure~\ref{fig:audience}). 

\begin{framed}
    \noindent \textbf{Takeaway \#2:} There is a lack of diversity in CS educator research, in terms of target educator, educational contexts and geographic locations, that research needs to carefully address.
\end{framed}

% \begin{figure}[htbp]
%     \centering
%     \includegraphics[width=\textwidth]{Images/orderedchallenges.png}
%     \caption{Number of Unique Paper References per CS Educator Challenge Category}
%     \label{fig:challenge-distribution}
% \end{figure}

% \vspace{0.5cm}

% \begin{figure}[htbp]
%     \centering
%     \includegraphics[width=\linewidth]{Images/course3.png}
%     \caption{Types of Courses Studied}
%     \label{fig:course}
% \end{figure}

\begin{figure}[htbp]
    \centering
    \includegraphics[width=0.95\textwidth]{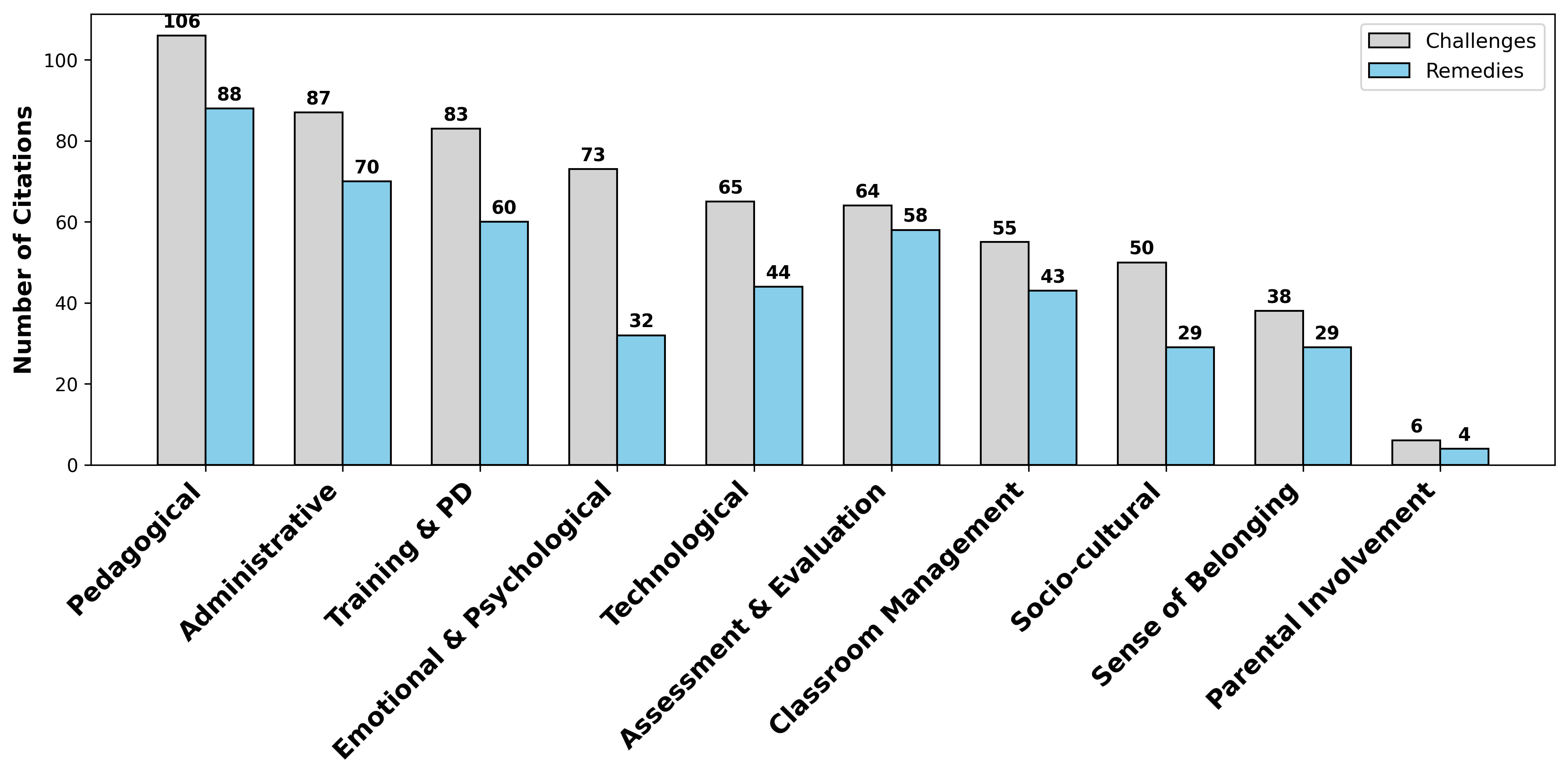}
    \caption{Comparison of Challenge and Remedy Citations Across CS Educator Challenge Categories}
    \label{fig:challenge-vs-remedy}
\end{figure}

\subsection{Key Educator Challenges and remedies (RQ1, RQ2)}
Our first research question, RQ1, focuses on the challenges CS educators face, and RQ2 are about remedies We found 108 papers mentioning educator challenges that fell into 10 major categories as summarized in Fig.~\ref{fig:challenge-vs-remedy}. We also summarize the remedies for these challenges alongside the challenges.

\subsubsection{Pedagogical Challenges}
CS educators face challenges effectively designing, delivering, and adapting instruction due to the evolving nature of computing~\cite{Ni2021,Vahid2024,Denny2024}, limited pedagogical content knowledge ~\cite{Thigpen2024,Musaeus2024} and inadequate training ~\cite{Venn-Wycherley2020,Goncharow2021,Tshukudu2021,Singh2020}. Preparation gaps, communication barriers~\cite{Farghally2024,Kirk2024} and challenges with specific pedagogical approaches such as project-based learning~\cite{Samarasekara2023,Samarasekara2023,Rosenbloom2023} are also reported. Curricular misalignment~\cite{Ni2021,Vahid2024,Denny2024}, ensuring industry relevance~\cite{Valstar2020,Williams2020,Pieper2020} and issues of catering to diverse student needs ~\cite{Sauerwein2023} or integrating with culturally-responsive pedagogy ~\cite{McGill2023,Hwang2023,Hwang2023} or open educational resources (OERs), especially in newer areas such as AI~\cite{Fowler2023,Lau2023} further make teaching a challenge.

\underline{Remedies:} Proportional to the number of papers highlighting pedagogical challenges, several papers also attempt to address these challenges via targeted remedies. To improve teaching effectiveness, instructors receive explicit pedagogical training in specific topics or materials~\cite{Yadav2022,McGill2024,Everson2022,Kumar2021}, while TAs and tutors receive training on feedback strategies, micro-teaching practice, and mentoring systems ~\cite{Cook2022,Zaman2023,Minnes2022,Akhmetov2024,Cheng2024,Riese2022,Riese2022}. Inclusive practices are integrated into training to create more welcoming classrooms \cite{Huang2023}. Programs such as the Make-a-Thon empower teachers to co-design AI curricula \cite{DiPaola2023}, and various curricular innovations—like contextualized learning\cite{Kumar2021,Venn-Wycherley2020,Samarasekara2023,Hamouda2025}, project-based instruction~\cite{Lee2021,Hooshangi2025,Repenning2021,Hamouda2025} and scaffolding~\cite{Vahid2024,Sauerwein2023}. Tailored strategies also address diverse learner needs, including disability support, bridging knowledge gaps, and adapting to generative AI tools~\cite{Yadav2022,Huff2021,Luu2023,Krause-Levy2022,Krause-Levy2023,Hooshangi2025,Mahon2024,Sheard2024,Lau2023}.

% Whereas the above relate to pedagogy and curricular content, teachers also face challenges related to administrative aspects, such as teacher shortage, lack of feedback and assessments and TA limitations~\cite{Forden2023,Forden2023,Zaman2023}. Specific areas such as debugging have also received attention~\cite{Macann2023}.

\subsubsection{Classroom Management Challenges}
Classroom management involves addressing logistical and resource constraints (e.g., time, devices, space)~\cite{Thigpen2024,Musaeus2024,Farghally2024,Kumar2022,Riese2022,Brunelle2022,Samarasekara2023,Forden2023}, ensuring student engagement~\cite{Venn-Wycherley2020,Vahid2024,Wang2021,Lau2023} and equitable participation ~\cite{Valstar2020,Goncharow2021,Pieper2020}. Managing conflict, peer help, varying levels of preparedness, and group dynamics in capstone projects are also mentioned~\cite{Hooshangi2025,Patel2024,Jia2022,Kao2020,Kussmaul2022,Krause-Levy2023,Riese2021}. TAs face challenges in coordination, high workload, time and student-queue management ~\cite{Minnes2022,Akhmetov2024,Kumar2022,Riese2022,Brunelle2022}, managing large classes and ensuring student wellbeing~\cite{Zaman2023,Huang2023,Lee2021,Huang2023}, while tutors are not always aligned with instructors~\cite{Brunelle2022,Malik2024}. Online learning adds additional challenges around student skills, integrity and visibility
~\cite{Markel2021,Rusak2021,Sentance2021,Goletti2021}.

\underline{Remedies:} To effectively manage classrooms amid growing class sizes, evolving technologies, and diverse learner needs, and educators leverage TAs and support teams~\cite{Mason2024,Akhmetov2024}, use automated tools for efficiency~\cite{Sauerwein2023,Wang2021,Forden2023,Vahid2024}, and restructure courses for scalability and self-contained learning~\cite{Brunelle2022,Kumar2022}. Research also addresses various issues surrounding classroom practices for student engagement, collaboration, participation and support ~\cite{Venn-Wycherley2020,Berry2022,Huang2023,Musaeus2024,Pieper2020,Vahid2024,Mahon2024,Nitta2022,Flatland2022,Skuratowicz2021,Lee2021,Huff2021}. Training also focuses on these aspects, along with cultural responsiveness practices~\cite{Rosenbloom2023,Hamouda2025,Yadav2022,Williams2020} for instructors as well as TAs and tutors.

\subsubsection{Assessment and Evaluation Challenges}
Educators struggle with creating timely, meaningful, outcome-aligned assessments ~\cite{Esche2024,Sauerwein2023,Denny2024,Luu2023,Mason2024}; making them adaptive, inclusive, equitable and scalable ~\cite{Tsan2022,Tsan2022b,Yadav2022,Lau2022}, supporting multiple formats ~\cite{Luxton-Reilly2023,Tsan2022,Su2021}, incorporating higher-order thinking and interdisciplinary work ~\cite{Izu2025,VanDeGrift2023} further complicate them. Ensuring academic integrity due to plagiarism, use of Internet and Generative AI ~\cite{Patel2024,Kirk2022,Brandes2020,Weintrop2022,Ju2020,Nitta2022} was another main challenge. Effective and transparent grading~\cite{Tsan2022,Wilde2024},
offering meaningful and personalized feedback~\cite{Esche2024,Upadhyaya2020,Kalathas2022,Nitta2022,Ishizue2024,Shrestha2022} and incorporating learning analytics and peer review~\cite{Marshall2020,Lee2022} present additional challenges. Online assessments add additional challenges around integrity, fair grading and collaborative work ~\cite{Cook2022,VanDeGrift2023,Krause-Levy2023,Saraiva2021,Hamouda2025}. 

\underline{Remedies:} To address challenges of assessments and feedback at scale, educators are leveraging automation and structured assessment practices (e.g., Automated Assessment Systems (APASs), ShellOnYou, TABot)~\cite{Berry2022,Forden2023,Sauerwein2023,Vahid2024}, autograders~\cite{Shanley2023},rubrics and exemplars~\cite{Esche2024,Kirk2022,Hooshangi2025},and automated code quality tools~\cite{Izu2025}. Unique and randomized exams~\cite{Rusak2021}, multiple assessment modalities~\cite{Mahon2024,Hamouda2025,Lau2023,Sheard2024},process-based assessment~\cite{Mahon2024,Sheard2024,Lau2023},plagiarism detection tools~\cite{Forden2023,Shrestha2022} and clear policies on AI usage~\cite{Mahon2024,Sheard2024} help address issues of integrity especially due to Generative AI use. TA training on academic integrity~\cite{Riese2022} and standardized assessments~\cite{Krause-Levy2023,Vahid2024}, and training educators on assessment principles and practices~\cite{Basu2022}, targeted diagnostic tools~\cite{McGill2024,Thigpen2024}, integrating assessment in course design~\cite{Farghally2024,Huang2023},calibration sessions for subjective grading~\cite{Kao2020,Hooshangi2025} and reframing assessment purpose~\cite{Basu2022} are also found to result in more robust, transparent, and equitable assessments.

\subsubsection{Emotional and Psychological Challenges}
CS educators report stress, burnout and low self-esteem~\cite{Thigpen2024,Farghally2024,Venn-Wycherley2020}, feelings of isolation ~\cite{Saraiva2021,Hamouda2025,Izu2025,Shrestha2022}, and fear of missing out~\cite{Marshall2020}, especially when adapting to new technologies ~\cite{Patel2024}. Emotional burdens also stem from assessments (e.g., offering feedback~\cite{Fowler2023,Forden2023,Lau2023}), providing emotional support to students~\cite{Yadav2022,Tsan2022b,Hamouda2025,Kalathas2022,Berry2022} and accommodating special requests~\cite{Huff2021}. TAs, underprepared, must deal with harrassment or emotionally-charged situations~\cite{Huang2023}. Lack of institutional and peer support exacerbate these issues ~\cite{Vivian2020,Goncharow2021,Tshukudu2021,Ni2021,Tsan2022}, as did the COVID-19 pandemic~\cite{Perlman2021,Skuratowicz2021,Markel2021}.

\underline{Remedies: }To reduce stress due to new technologies, targeted professional development (PD) and training~\cite{Vivian2020,Zaman2023} as well as teacher involvement in them in co creating courses~\cite{Minnes2022,Goletti2021,Farghally2024} have been explored. Community building to foster a sense of belonging~\cite{Flatland2022,Kussmaul2022,Perlman2021,Jocius2021,Skuratowicz2021,Tsan2022,Lee2022,Wilde2024,Haglund2024}, facilitate collaborative learning~\cite{DiPaola2023,Musaeus2024,Lehtimki2022,Hamouda2025} and increased instructor support feedback and support 
~\cite{Kirk2022,Cook2022,Zaman2023,Wang2021} have also been explored. Studies also address aspects such as creating safe and inclusive environments~\cite{Huang2023,Krause-Levy2023}, destigmatizing mistakes~\cite{Macann2023,Berry2022}, flexibility and adaptability in programs~\cite{Tsan2022,Mahoney2023} and direct support for mental health and difficult topics ~\cite{Farghally2024,Riese2022,Minnes2022,Hu2023,Parthasarathy2024}.

\subsubsection{Technological Challenges}
Two major themes emerge here. First, educators must deal with issues of technical infrastructure or access for their learning and teaching (e.g., unavailability or outdated devices, internet, incompatible software, licenses) ~\cite{Cook2022,Saraiva2021,Patel2024,Izu2025,Musaeus2024,Kalathas2022,Nitta2022,Ishizue2024,Tsan2022}. Second, they must deal with their own lack of technical knowledge ~\cite{Marshall2020,Kirk2022,Lee2022}, digital and tool literacy~\cite{Farghally2024,Shrestha2022,Brandes2020,VanDeGrift2023,Mason2024,Berry2022}, or having to cope with rapidly evolving topics such as AI or cybersecurity to align with industry needs~\cite{Valstar2020,Denny2024}. Difficulties handling and recovering from various errors issues~\cite{Vivian2020,Pieper2020}, lack of prompt technical support when stuck~\cite{Weintrop2022,Luxton-Reilly2023} and poor usability or interoperability of tools worsen teachers' technical challenges~\cite{Wang2021,Tshukudu2021,McGill2023,Fowler2023}. The COVID pandemic exposed several of these gaps~\cite{Perlman2021,Skuratowicz2021,Simmonds2021}, calling for concerted efforts to address them. 

\underline{Remedies:} These technological issues have primarily been addressed by providing teachers with resources that are user-friendly, offer scaffolding and minimize errors. Examples are web-Based and containerized environments~\cite{Kumar2022,Nitta2022,Forden2023}, low-cost devices and open-source tools~\cite{Hamouda2025}, user-friendly interfaces~\cite{Wang2021,Forden2023} and living and centralized repositories~\cite{Grover2024,Blanchard2022,Fowler2023} of teacher resources. To address issues of limited technical infrastructure to teachers and studies, inclusive adaptation of the material, such as low-tech, unplugged, low-bandwidth or offline alternatives are explored ~\cite{Lehtimki2022,Malik2024,Flatland2022,Hamouda2025,Macann2023,Grover2024}. Technical support, institutional roles~\cite{Jocius2021,Skuratowicz2021,Flatland2022}, pedagogical and technical training~\cite{Riese2021,Riese2022,Williams2020}, automation of workflows~\cite{Akhmetov2024} and readjusting curricula and pedagogical tools to work with and adopt newer technologies such as AI ~\cite{Ishizue2024,Denny2024,Lau2023,Cucuiat2024} have also shown to help with technical challenges for educators.

\subsubsection{Administrative and Institutional Challenges}
Systemic issues within institutions such as rigid and outdated curricula, ambiguous role definitions, resource shortages, and imbalanced workloads often restrict effective teaching \cite{Farghally2024,Valstar2020,Williams2020,Hill2021,Luxton-Reilly2023}. Lack of organizational support and incentives for professional development and training programmes  ~\cite{Zaman2023,Huang2023,Minnes2022,Riese2022,Akhmetov2024,McGill2023,Hu2023,Greifenstein2023,Lau2023}, lack of standards and guidelines around sharing and using open educational resources \cite{Hwang2023,Rosenbloom2023,Fowler2023,Repenning2021,Goncharow2021,Ni2021,Jia2022,Marshall2020} and outdated/underspecified rules and policy guidelines surrounding areas such as instructional quality\cite{Ju2020,Upadhyaya2020,Kalathas2022,Krause-Levy2023}, online teaching~\cite{Kirk2022,Tsan2022,Nitta2022,Shrestha2022}, interdepartmental collaborations, workload distribution or content ownership \cite{Lau2022} are also reported. Issues of budget, accreditation and bureaucracy~\cite{Ishizue2024,Mason2024,Tsan2022} also present hurdles in teaching ~\cite{Weintrop2022,Brandes2020,Shrestha2022}. 

\underline{Remedies:} In terms of solutions, papers propose ways of addressing staffing shortages and balance workload such as hiring adjuncts and alumni~\cite{VanDeGrift2023}, expanding recruitment pipelines~\cite{Weintrop2022}, tiered TA Structures~\cite{Akhmetov2024}, use of automated tools~\cite{Sauerwein2023,Forden2023,Wang2021,Kumar2022}, centralized course materials and structures~\cite{Vahid2024}, and offering financial incentives~\cite{Everson2022,Yadav2022,Perlman2021}. Improving the efficiency of existing teachers 
 Enhanced teacher preparation and professional development via formal certification pathways~\cite{Yadav2022,Flatland2022}, scalable professional development models~\cite{Jetzinger2024,Tsan2022}, and long-term institutional investment~\cite{Yadav2022,Flatland2022,Huff2021,Saraiva2021,Hamouda2025} and strategic alliances~\cite{Simmonds2021,Hamouda2025,Friend2022}for teacher career development are also noted. Finally, policy around standardized curricula~\cite{Upadhyaya2020,Jia2022,Vahid2024}, stakeholder education~\cite{Grover2024,Simmonds2021,Parthasarathy2024}, flexibility ~\cite{Jia2022,Hwang2023,Brunelle2022,Skuratowicz2021}, time constraints, recognition for faculty contributions~\cite{Fowler2023} and use of tools such as AI~\cite{Mahon2024,Lau2023} have been proposed.

\subsubsection{Training and Professional Development Challenges}
Teachers lack pedagogical preparation and equity training~\cite{Krause-Levy2022,Akhmetov2024,Kalathas2022,Ishizue2024,Mason2024}. At K-12 level, many educators also lack prior coding experience, struggle with computational thinking integration into disciplinary teaching, and require sustained scaffolding and exposure~\cite{Jocius2021}. Most existing professional development programs frequently fail to differentiate between educator roles (e.g., tutors, TAs, and classroom teachers)~\cite{Hwang2023,Yadav2022Review,Parthasarathy2024,Friend2022}, are often narrow in scope and for short durations~\cite{Ju2020,Luxton-Reilly2023}, and are inaccessible due to various constraints~\cite{Sentance2021,Thigpen2024,Nugent2022,Grover2024}. Teachers need sustained mentoring\cite{Ju2020,Yadav2022}, hands-on digital or subject specific training \cite{Anyango2021,Saraiva2021}, support for emerging tools like AI and VR~\cite{Tsan2022,Wilde2024}, but lack of flexibility \cite{Cook2022}, misaligned incentives \cite{Krause-Levy2023}, unclear and unmeasured long-term impact of training \cite{Shrestha2022,VanDeGrift2023} and integrating new tools and pedagogies ~\cite{Upadhyaya2020} further devalue professional development programs. 

\underline{Remedies:} Attempts at improving training and professional development emphasize on making them long-term, iterative~\cite{Tsan2022,Lee2022,DiPaola2023}, focused on specific topics\cite{Kumar2021,Everson2022,Flatland2022,Kumar2021,McGill2024,Grover2024,Hu2023,Marshall2020,Yadav2022} or pedagogical techniques and issues~\cite{Izu2025,Tshukudu2021,Munasinghe2021,Parthasarathy2024,Hu2023}, as against the traditional short and generic ones. Good professional development programs also integrate content and pedagogy~\cite{Basu2022,Everson2022,Tsan2022,Yadav2022,Macann2023,Jia2022,Jetzinger2024}, foster hands-on and active learning, ~\cite{Patel2024,Tsan2022,Skuratowicz2021,Hamouda2025,DiPaola2023,Musaeus2024,Lehtimki2022,Grover2024} and offer mentoring and peer support ~\cite{Kussmaul2022,Minnes2022,Flatland2022,Wilde2024,Haglund2024,Markel2021,Riese2021}. Research also recommends making them flexible and scalable~\cite{Jetzinger2024,Tsan2022,Perlman2021,Minnes2022,Zaman2023,Riese2022}, offering just-in-time support for TAs right before tutoring sessions \cite{Cheng2024}, and incorporating technical assistance, resources, incentives and documentation ~\cite{McGill2023,Jocius2021,Skuratowicz2021,Everson2022,Flatland2022,Wilde2024,McGill2024,Zaman2023,Kussmaul2022}.

\subsubsection{Socio-cultural Challenges}
Educators face underrepresentation, cultural mismatch, and DEI resistance~\cite{Hwang2023,Everson2022,Rosenbloom2023}. Educational materials often fail to reflect diverse realities~\cite{Greifenstein2023,Simmonds2021,Kalathas2022} and linguistic backgrounds \cite{Nitta2022}. International educators experience additional challenges around language and culture~\cite{Markel2021,Riese2021,Minnes2022,Krause-Levy2022}. Issues around AI algorithm biases and disempowerment~\cite{Lau2023,Sheard2024,Mahon2024}, digital divide and inequitable access to resources or devices ~\cite{Luxton-Reilly2023,Upadhyaya2020,Luxton-Reilly2023,Lau2022} and gender, generational and cultural gaps in tech use~\cite{Krause-Levy2023,Saraiva2021,Hamouda2025} highlight the need for equitable and culturally-sensitive and inclusive pedagogical content for classrooms--both for the benefit of teachers and students.

\underline{Remedies:} Addressing DEI-related challenges are primarily through culturally responsive education \cite{Hill2021,Venn-Wycherley2020,Hwang2023,Hamouda2025,Wilde2024,Hamouda2025}, training to address cultural aspects, bias and stereotypes\cite{Huang2023,Simmonds2021,Zaman2023}, broadening participation\cite{Hill2021,Flatland2022,Nugent2022,Weintrop2022} and explicit integration of ethics and social topics in education \cite{Grover2024,Lee2022,Mahon2024,Hu2023,Rosenbloom2023,Yadav2022}. Fostering inclusive learning\cite{DiPaola2023,Perlman2021,Flatland2022,Malik2024}, building communities, collaboration and peer support \cite{Hamouda2025,Sentance2021}and addressing individualized and contextual needs \cite{Venn-Wycherley2020,Hwang2023,Hamouda2025} are also explored.

\subsubsection{Parental and Community Involvement Challenges}
There are also several challenges relating to the parents and their involvement in CS education, especially at the K-12 level. Areas such as CS and AI education require raising public awareness and obtaining "buy-in" from various key educational stakeholders. Depending on neighborhoods, teachers also have the added responsiblity of "assisting families help their children do well in school" ~\cite{Jia2022}, or deal with parental skepticism especially when the latter are tech-educated ~\cite{Everson2022}. Compounded by a lack of broader support for teachers from institution or peers, teachers face low self-efficacy~\cite{Vivian2020} and struggle with cultural norms and their conflict with various group activities~\cite{Hamouda2025}.

\underline{Remedies:} To foster a sense of belonging, especially for K-12 teachers, who often work in isolation, building professional communities for support, growth and collaboration have been a large focus in literature\cite{Flatland2022,Kussmaul2022,Perlman2021,Ni2021,Hill2021,Tsan2022,Lee2022,Nugent2022,Yadav2022,Wilde2024}. These include regular meetings\cite{Flatland2022,Marshall2020,Nugent2022,Wilde2024,Jetzinger2024},shared online spaces (e.g., Slack, discussion fora)\cite{Jocius2021,Nugent2022,Lee2022}, collaborative activities \cite{DiPaola2023,Lehtimki2022,Musaeus2024}, mentorship and peer Support\cite{Markel2021,Riese2021,Riese2022,Wilde2024},and structured training programs \cite{Riese2022,Minnes2022,Akhmetov2024,Haglund2024,Huang2023}. Encouraging participations in teacher organizations and conferences (e.g., CSTA, CSPathshala and CTiS in India)also help educators expand their networks and feel a sense of belonging \cite{Flatland2022,Ni2021}. Institutional support and recognition, as well as policy around collaborations are also discussed in this regard \cite{Vahid2024,Fowler2023}.

\subsubsection{Sense of Community/Belonging Challenges}
Finally, a recurring barrier in CS education is the isolation and lack of peer connection experienced by many educators, especially those who serve as the sole CS teacher in their institutions. Lack of incentives, no facilitation for collaboration (e.g., for OERss~\cite{Fowler2023,Blanchard2022}, in learning communities ~\cite{Flatland2022,Parthasarathy2024}) or extant community-building and mentorship efforts ~\cite{Saraiva2021,Izu2025} often underlie such challenges, resulting in limited collaboration~\cite{Ni2021,Yadav2022,Nugent2022}, lack of peer support networks or professional development~\cite{Skuratowicz2021,Basu2022}. Challenges also arise from fragmented or distributed TA teams, or remote formats of learning and working ~\cite{Akhmetov2024,Riese2022,Minnes2022,Haglund2024,Lehtimki2022}. Cultural norms and gender bias further lower belonging~\cite{Rosenbloom2023,Everson2022,Hwang2023}.

\underline{Remedies:} A small number of papers, especially in the context of Western K-12 education, address raising awareness about CS and AI~\cite{DiPaola2023,Samarasekara2023,Yadav2022}, enabling broader understanding and acceptance of CS as a viable academic and career path~\cite{Williams2020}, fostering collaborations with industry\cite{Hooshangi2025,Lee2021,Parthasarathy2024}, community \cite{Rosenbloom2023,Yadav2022,Hwang2023}, alumni~\cite{VanDeGrift2023}, volunteers~\cite{Williams2020} and community and school-college partnerships\cite{Flatland2022,Hill2021}, to improve education. Engaging parents via home-based resources and exercises \cite{Lehtimki2022} and inviting parents to address issues of culture are also explored~\cite{Hamouda2025}.

\begin{framed}
    \noindent \textbf{Takeaway \#3:} Educator challenges fall into ten major categories: pedagogical, classroom management, assessment and evaluation, emotional and psychological, technological, administrative and institutional, socio-cultural, parental and sense of community. Researchers have also proposed various remedial solutions for these challenges; see Figure~\ref{fig:challenge-vs-remedy} for their relative occurrence in literature.
\end{framed}

\subsection{Gaps and open opportunities (RQ3)}

Our synthesis of 106 relevant studies revealed an imbalance in the attention given to various categories of challenges faced by CS educators (Figure~\ref{fig:challenge-vs-remedy}). First, while pedagogical, emotional and psychological, technological, administrative, and professional development challenges have received relatively greater attention in terms of problem characterization, only pedagogical, administrative and teacher training aspects have received proportionate remedial attempts. Teachers' technical challenges and problems of well-being have been well described, but not much has been attempted by way of alleviating these challenges--resulting in persistent challenges. Assessment and evaluation is another area that has received attention, but lesser compared to pedagogical aspects, even though development of meaningful assessments in the day and age of Generative AI is emerging to be critical.

Other areas remain under-addressed in literature; these are classroom management, socio-cultural aspects, and fostering a sense of community and belonging among educators and receiving community support. Any works in these areas are also limited to K-12 levels; for example, socio-cultural considerations are largely limited to contextualization and unplugged approaches in K-12 levels, as does parental involvement and outreach to assist with education. 

These overlooked categories are important, because they can have cascading effects on other challenge domains. For instance, inadequate classroom management strategies can hinder pedagogical innovation, and limited parental or community engagement can diminish the efficacy of student motivation and outreach efforts. Similarly, neglecting socio-cultural and belonging issues can impede inclusion, retention, and student trust, thus undermining broader educational reform goals. Thus,

\begin{framed}
    \noindent \textbf{Takeaway \#4:} The imbalance in focus towards various educator challenges and their remedies result in persistence of these challenges and presents opportunities for future research.
\end{framed}

The above results must also be viewed in the light of the broad trends we described at the start of the section. Notably, there is a general lack of diversity in educator-focused research--with most research focused in US and Canada, UK, EU and Australia. Research emerging from the rest of the world--including India, Africa and other Asian nations are still sparse. Considering the hyper-local contexts in which education is often situated, this lack of diversity in educator research highlights the need for understanding and addressing them in renewed vigor, in specific contexts. 

Specifically concerning, the Indian CS education research community, only two papers in top CS education venues in the last five years were from an Indian study setting. A careful review of proceedings of conferences such as ACM COMPUTE or CTiS (the annual ACM CSPathshala event for computing school teachers in India) revealed no papers focusing on educator issues--although there have been keynote talks in the recent years focusing on the issues. These findings highlight a broader issue in computing research--worldwide, but more so in India--where technical and pedagogy-centred research overshadow human-centred needs, while the latter might in fact hold the key to several barriers to ensuring effective CS education.

\begin{framed}
    \noindent \textbf{Takeaway \#5:} In an Indian context, the lack of India-centric educator studies in major CS education conferences in India and worldwide highlight an overall pedagogy-centricity as against teacher-centricity--calling for revisiting research priorities.
\end{framed}

\section{Discussion: A call for action}

Our systematic mapping study reveals a complex and multi-dimensional landscape of challenges faced by CS educators. While pedagogical, emotional, and institutional challenges have received notable attention, several domains critical to teaching effectiveness remain underexplored. For instance, challenges related to classroom management, socio-cultural barriers, and educators' sense of belonging are essential for fostering inclusive and responsive classrooms, yet these topics lack sufficient empirical grounding and practical solutions in the literature.

The uneven distribution of scholarly focus reflects a broader issue in computing education research, where structural and curriculum-centered reforms often overshadow human-centered needs. TAs and tutors—who serve as critical instructional intermediaries—frequently face coordination breakdowns, emotional strain, and training gaps, which are insufficiently addressed in both research and institutional policy. Similarly, community engagement and parental involvement are rarely discussed, though they play a crucial role in broadening participation and supporting student success in K–12 and pre-service education settings. 

Notably, there is a lack of India-specific studies, which hurts the development of CS education and educators in the country. With hierarchy in schools, tiering of colleges, balkanization in terms of linguistic, cultural and curricular diversity and widespread inequities, CS educator issues need urgent attention in the Indian context--not only limited to the 10 categories of challenges listed earlier, but also to unearth newer local ones that need local solutions. 

There are also opportunities for learning from solutions elsewhere. For example, there are opportunities for community development, school-college or college-college partnerships as is done in community-building in the US. While such efforts do exist in India (ACM India teaching partnership programs, CSPathshala communities), rigorous research on their effectiveness, learnings and the socio-technical aspects are amiss.

\section{Conclusion}

With the goal of gaining a holistic understanding of educator issues in CS education, this study presents a systematic mapping of the challenges faced by CS educators and the interventions proposed to address them, based on a review of 108 peer-reviewed papers published over the last five years. We identified ten major categories of challenges--with uneven distribution in focus received in research, both in terms of understanding the challenges as well as in solutions. Notably, there is an overfocus on pedagogical, administrative and evaluation approaches, while human-centric aspects such as technical challenges or community-building remain understudied. Our paper aims to foreground these overlooked yet crucial dimensions of CS educator experience by synthesizing existing literature and identifying significant gaps that warrant further inquiry.

We also highlight the urgent need for localized research in diverse settings like India, where systemic inequities and cultural heterogeneity pose unique challenges. By surfacing underexplored problem areas and pointing toward potential solution pathways—both indigenous and globally informed—this paper lays the groundwork for more inclusive, context-sensitive, and sustainable improvements in CS educator support and development. Thus, through our systematic mapping study, we advocate for a more holistic and human-centered research agenda in computing education research.

\newpage

% \section{Future Work}

\bibliographystyle{splncs04}
\bibliography{export}

\begin{thebibliography}{100}
\providecommand{\url}[1]{\texttt{#1}}
\providecommand{\urlprefix}{URL }
\providecommand{\doi}[1]{https://doi.org/#1}

\bibitem{Akhmetov2024}
Akhmetov, I., Ahmed, S., Ayuno, K.: How we manage an army of teaching assistants: Experience report on scaling a cs1 course. In: Proc. of the 55th SIGCSE TS V. 1. pp. 32--38. ACM (2024). \doi{10.1145/3626252.3630871}

\bibitem{Anyango2021}
Anyango, J.T., Suleman, H.: Supporting cs1 instructors: Design and evaluation of a game generator. In: Proc. of the 26th ITiCSE V. 1. pp. 115--121. ACM (2021). \doi{10.1145/3430665.3456306}

\bibitem{Bassner2024}
Bassner, P., Frankford, E., Krusche, S.: Iris: An ai-driven virtual tutor for computer science education. In: Proc. of the 2024 on Innovation and Technology in Computer Science Education V. 1. pp. 394--400. ACM (2024). \doi{10.1145/3649217.3653543}

\bibitem{Basu2022}
Basu, S., Rutstein, D., Tate, C., Rachmatullah, A., Yang, H.: Standards-aligned instructional supports to promote computer science teachers' pedagogical content knowledge. In: Proc. of the 53rd SIGCSE TS. pp. 404--410. ACM (2022). \doi{10.1145/3478431.3499403}

\bibitem{Berry2022}
Berry, V., Castelltort, A., Pelissier, C., Rousseau, M., Tibermacine, C.: Shellonyou. In: Proc. of the 27th ACM Conference on on Innovation and Technology in Computer Science Education Vol. 1. pp. 379--385. ACM (2022). \doi{10.1145/3502718.3524753}

\bibitem{Blanchard2022}
Blanchard, J., Hott, J.R., Berry, V., Carroll, R., Edmison, B., Glassey, R., Karnalim, O., Plancher, B., Russell, S.: Stop reinventing the wheel! promoting community software in computing education. In: Proc. of the 2022 Working Group Reports on Innovation and Technology in Computer Science Education. pp. 261--292. ACM (2022). \doi{10.1145/3571785.3574129}

\bibitem{Brandes2020}
Brandes, O., Armoni, M.: Towards a holistic reservoir of research-based pck segments of k-12 computer science teachers. In: Proc. of the 2020 ITiCSE. pp. 131--137. ACM (2020). \doi{10.1145/3341525.3387412}

\bibitem{Brunelle2022}
Brunelle, N., Evans, D.: Comfortable cohorts and tractable teams. In: Proc. of the 53rd SIGCSE TS. pp. 717--723. ACM (2022). \doi{10.1145/3478431.3499353}

\bibitem{Budgen2008Mapping}
Budgen, D., Turner, M., Brereton, P., Kitchenham, B.: Using mapping studies in software engineering. Tech. rep., EBSE (2008), \url{http://www.ebse.org.uk/}

\bibitem{Cheng2024}
Cheng, A.Y., Tanimura, E., Tey, J., Wu, A.C., Brunskill, E.: Brief, just-in-time teaching tips to support computer science tutors. In: Proc. of the 55th SIGCSE TS V. 1. pp. 200--206. ACM (2024). \doi{10.1145/3626252.3630794}

\bibitem{Cook2022}
Cook, A., Phan, V., Windsor, A.: Improving ta feedback on in-class coding assignments for introductory computer science. In: Proc. of the 27th ACM Conference on on Innovation and Technology in Computer Science Education Vol. 1. pp. 421--427. ACM (2022). \doi{10.1145/3502718.3524746}

\bibitem{Cucuiat2024}
Cucuiat, V., Waite, J.: Feedback literacy: Holistic analysis of secondary educators' views of llm explanations of program error messages. In: Proc. of the 2024 on Innovation and Technology in Computer Science Education V. 1. pp. 192--198. ACM (2024). \doi{10.1145/3649217.3653595}

\bibitem{Denny2024}
Denny, P., MacNeil, S., Savelka, J., Porter, L., Luxton-Reilly, A.: Desirable characteristics for ai teaching assistants in programming education. In: Proc. of the 2024 on Innovation and Technology in Computer Science Education V. 1. pp. 408--414. ACM (2024). \doi{10.1145/3649217.3653574}

\bibitem{DiPaola2023}
DiPaola, D., Moore, K.S., Ali, S., Perret, B., Zhou, X., Zhang, H., Lee, I.: Make-a-thon for middle school ai educators. In: Proc. of the 54th SIGCSE TS V. 1. pp. 305--311. ACM (2023). \doi{10.1145/3545945.3569743}

\bibitem{Esche2024}
Esche, S.: Rubric for the quality of answers to student queries about code. In: Proc. of the 55th SIGCSE TS V. 1. pp. 331--337. ACM (2024). \doi{10.1145/3626252.3630918}

\bibitem{Everson2022}
Everson, J., Ko, A.J.: “i would be afraid to be a bad cs teacher”: Factors influencing participation in pre-service secondary cs teacher education. In: Proc. of the 2022 ACM Conference on International Computing Education Research - Volume 1. pp. 237--246. ACM (2022). \doi{10.1145/3501385.3543966}

\bibitem{Farghally2024}
Farghally, M., Seyam, M., Shaffer, C.A.: Towards establishing a training program to support future cs teaching-focused faculty. In: Proc. of the 55th SIGCSE TS V. 1. pp. 338--344. ACM (2024). \doi{10.1145/3626252.3630798}

\bibitem{Flatland2022}
Flatland, R., Matthews, J., White, P., Egan, M., Moya, J.: Building cs teacher capacity through comprehensive college/high school partnerships. In: Proc. of the 53rd SIGCSE TS. pp. 606--612. ACM (2022). \doi{10.1145/3478431.3499364}

\bibitem{Forden2023}
Forden, J., Gebhard, A., Brylow, D.: Experiences with ta-bot in cs1. In: Proc. of the ACM Conference on Global Computing Education Vol 1. pp. 57--63. ACM (2023). \doi{10.1145/3576882.3617930}

\bibitem{Fowler2023}
Fowler, M., IV, D.H.S., Chen, B., Zilles, C.: "\"i don't gamble to make my livelihood\": Understanding the incentives for. In: Proc. of the 2023 ACM Conference on International Computing Education Research V.1. pp. 430--443. ACM (2023). \doi{10.1145/3568813.3600136}

\bibitem{Friend2022}
Friend, M., Twarek, B., Koontz, J., Bell, A., Joseph, A.: Trends in cs teacher professional development. In: Proc. of the 53rd SIGCSE TS. pp. 390--396. ACM (2022). \doi{10.1145/3478431.3499292}

\bibitem{Goletti2021}
Goletti, O., Mens, K., Hermans, F.: Tutors' experiences in using explicit strategies in a problem-based learning introductory programming course. In: Proc. of the 26th ITiCSE V. 1. pp. 157--163. ACM (2021). \doi{10.1145/3430665.3456348}

\bibitem{Goncharow2021}
Goncharow, A., Mcquaigue, M., Saule, E., Subramanian, K., Payton, J., Goolkasian, P.: Mapping materials to curriculum standards for design, alignment, audit, and search. In: Proc. of the 52nd SIGCSE TS. pp. 295--301. ACM (2021). \doi{10.1145/3408877.3432388}

\bibitem{Greifenstein2023}
Greifenstein, L., Heuer, U., Fraser, G.: Exploring programming task creation of primary school teachers in training. In: Proc. of the 2023 Conference on Innovation and Technology in Computer Science Education V. 1. pp. 471--477. ACM (2023). \doi{10.1145/3587102.3588809}

\bibitem{Grover2024}
Grover, S.: Teaching ai to k-12 learners: Lessons, issues, and guidance. In: Proc. of the 55th SIGCSE TS V. 1. pp. 422--428. ACM (2024). \doi{10.1145/3626252.3630937}

\bibitem{Haglund2024}
Haglund, P., Mannila, L., Strömbäck, F., Berglund, A.: Grasping the unseen: Ta insights into teaching subtle concepts in computer science. In: Proc. of the 2024 on Innovation and Technology in Computer Science Education V. 1. pp. 157--163. ACM (2024). \doi{10.1145/3649217.3653601}

\bibitem{Hamouda2025}
Hamouda, S., Marshall, L., Sanders, K., Tshukudu, E., Adelakun-Adeyemo, O., Becker, B.A., Dodoo, E.R., Korsah, G.A., Luvhengo, S., Ola, O., Parkinson, J., Sanusi, I.T.: Computing education in african countries: A literature review and contextualised learning materials. In: 2024 Working Group Reports on Innovation and Technology in Computer Science Education. pp. 1--33. ACM (2025). \doi{10.1145/3689187.3709606}

\bibitem{Hill2021}
Hill, K., Fancsali, C.: Bridging professional development to practice: Using school support visits to build teacher confidence in delivering equitable cs instruction. In: Proc. of the 52nd SIGCSE TS. pp. 725--731. ACM (2021). \doi{10.1145/3408877.3432414}

\bibitem{Hooshangi2025}
Hooshangi, S., Shakil, A., Dasgupta, S., Davis, K.C.C., Farghally, M., Fitzpatrick, K., Gutica, M., Hardt, R., Riddle, S., Seyam, M.: Instructors' perspectives on capstone courses in computing fields: A mixed-methods study. In: 2024 Working Group Reports on Innovation and Technology in Computer Science Education. pp. 68--94. ACM (2025). \doi{10.1145/3689187.3709608}

\bibitem{Hu2023}
Hu, A.D., Yadav, A.: How k-12 cs teachers conceptualize cs ethics. In: Proc. of the 54th SIGCSE TS V. 1. pp. 910--916. ACM (2023). \doi{10.1145/3545945.3569775}

\bibitem{Huang2023}
Huang, V., Fox, A.: A climate-first approach to training student teaching assistants. In: Proc. of the 54th SIGCSE TS V. 1. pp. 423--429. ACM (2023). \doi{10.1145/3545945.3569826}

\bibitem{Huff2021}
Huff, E.W., Boateng, K., Moster, M., Rodeghero, P., Brinkley, J.: Exploring the perspectives of teachers of the visually impaired regarding accessible k12 computing education. In: Proc. of the 52nd SIGCSE TS. pp. 156--162. ACM (2021). \doi{10.1145/3408877.3432418}

\bibitem{Hwang2023}
Hwang, Y., Das, A., Waite, J., Sentance, S.: Using a sociological lens to investigate computing teachers’ culturally responsive classroom practices. In: Proc. of the 2023 ACM Conference on International Computing Education Research V.1. pp. 206--221. ACM (2023). \doi{10.1145/3568813.3600112}

\bibitem{Ishizue2024}
Ishizue, R., Sakamoto, K., Washizaki, H., Fukazawa, Y.: Improved program repair methods using refactoring with gpt models. In: Proc. of the 55th SIGCSE TS V. 1. pp. 569--575. ACM (2024). \doi{10.1145/3626252.3630875}

\bibitem{Izu2025}
Izu, C., Mirolo, C., Börstler, J., Connamacher, H., Crosby, R., Glassey, R., Haldeman, G., Kiljunen, O., Kumar, A.N., Liu, D., Luxton-Reilly, A., Matsumoto, S., de~Oliveira, E.C., Russell, S., Shah, A.: Introducing code quality at cs1 level: Examples and activities. In: 2024 Working Group Reports on Innovation and Technology in Computer Science Education. pp. 339--377. ACM (2025). \doi{10.1145/3689187.3709615}

\bibitem{Jetzinger2024}
Jetzinger, F., Baumer, S., Michaeli, T.: Artificial intelligence in compulsory k-12 computer science classrooms: A scalable professional development offer for computer science teachers. In: Proc. of the 55th SIGCSE TS V. 1. pp. 590--596. ACM (2024). \doi{10.1145/3626252.3630782}

\bibitem{Jia2022}
Jia, X., Hermans, F.: Teaching quality in programming education:. In: Proc. of the 2022 ACM Conference on International Computing Education Research - Volume 1. pp. 223--236. ACM (2022). \doi{10.1145/3501385.3543962}

\bibitem{Jocius2021}
Jocius, R., Joshi, D., Albert, J., Barnes, T., Robinson, R., Cateté, V., Dong, Y., Blanton, M., O'Byrne, I., Andrews, A.: The virtual pivot. In: Proc. of the 52nd SIGCSE TS. pp. 1198--1204. ACM (2021). \doi{10.1145/3408877.3432558}

\bibitem{Ju2020}
Ju, A., Hemani, A., Dimitriadis, Y., Fox, A.: What agile processes should we use in software engineering course projects? In: Proc. of the 51st SIGCSE TS. pp. 643--649. ACM (2020). \doi{10.1145/3328778.3366864}

\bibitem{Kalathas2022}
Kalathas, P., Parham-Mocello, J., Elliot, R., Lockwood, E.: Exploring math + cs in a secondary education methods course. In: Proc. of the 53rd SIGCSE TS. pp. 689--695. ACM (2022). \doi{10.1145/3478431.3499405}

\bibitem{Kao2020}
Kao, Y., Nolan, I., Rothman, A.: Project scoring for program evaluation and teacher professional development. In: Proc. of the 51st SIGCSE TS. pp. 1133--1138. ACM (2020). \doi{10.1145/3328778.3366959}

\bibitem{Kirk2022}
Kirk, D., Crow, T., Luxton-Reilly, A., Tempero, E.: Teaching code quality in high school programming courses - understanding teachers’ needs. In: Proc. of the 24th Australasian Computing Education Conference. pp. 36--45. ACM (2022). \doi{10.1145/3511861.3511866}

\bibitem{Kirk2024}
Kirk, D., Luxton-Reilly, A., Tempero, E.: A literature-informed model for code style principles to support teachers of text-based programming. In: Proc. of the 26th Australasian Computing Education Conference. pp. 134--143. ACM (2024). \doi{10.1145/3636243.3636258}

\bibitem{Krause-Levy2022}
Krause-Levy, S., Lim, R.S., Molina, I.V., Cao, Y., Porter, L.: An exploration of student-tutor interactions in computing. In: Proc. of the 27th ACM Conference on on Innovation and Technology in Computer Science Education Vol. 1. pp. 435--441. ACM (2022). \doi{10.1145/3502718.3524786}

\bibitem{Krause-Levy2023}
Krause-Levy, S., Salguero, A., Lim, R.S., McTavish, H., Trajkovic, J., Porter, L., Griswold, W.G.: Instructor perspectives on prerequisite courses in computing. In: Proc. of the 54th SIGCSE TS V. 1. pp. 277--283. ACM (2023). \doi{10.1145/3545945.3569787}

\bibitem{Kumar2022}
Kumar, A.N.: Solvelets. In: Proc. of the 27th ACM Conference on on Innovation and Technology in Computer Science Education Vol. 1. pp. 151--157. ACM (2022). \doi{10.1145/3502718.3524811}

\bibitem{Kumar2021}
Kumar, V., Karkare, A.: Instructor performance on progressively complex programming tasks: A multi-institutional study from india. In: Proc. of the 26th ITiCSE V. 1. pp. 561--567. ACM (2021). \doi{10.1145/3430665.3456384}

\bibitem{Kussmaul2022}
Kussmaul, C., Hu, H.H., Campbell, P.B., Mayfield, C., Yadav, A.: Professional development and support for pogil in computer science. In: Proc. of the 53rd SIGCSE TS. pp. 738--744. ACM (2022). \doi{10.1145/3478431.3499381}

\bibitem{Lau2023}
Lau, S., Guo, P.: From "ban it till we understand it" to "resistance is futile": How university programming instructors plan to adapt as more students use ai code generation and explanation tools such as chatgpt and github copilot. In: Proc. of the 2023 ACM Conference on International Computing Education Research V.1. pp. 106--121. ACM (2023). \doi{10.1145/3568813.3600138}

\bibitem{Lau2022}
Lau, S., Nolan, D., Gonzalez, J., Guo, P.J.: How computer science and statistics instructors approach data science pedagogy differently. In: Proc. of the 53rd SIGCSE TS. pp. 29--35. ACM (2022). \doi{10.1145/3478431.3499384}

\bibitem{Lee2021}
Lee, E.S.A., Kuber, K., Rohian, H., Woodhead, S.: Pillars of program design and delivery: A case study using self-directed, problem-based, and supportive learning. In: Proc. of the 52nd SIGCSE TS. pp. 205--211. ACM (2021). \doi{10.1145/3408877.3432458}

\bibitem{Lee2022}
Lee, I., Zhang, H., Moore, K., Zhou, X., Perret, B., Cheng, Y., Zheng, R., Pu, G.: Ai book club. In: Proc. of the 53rd SIGCSE TS. pp. 202--208. ACM (2022). \doi{10.1145/3478431.3499318}

\bibitem{Lehtimki2022}
Lehtimäki, T., Monahan, R., Mooney, A., Casey, K., Naughton, T.J.: Bebras-inspired computational thinking primary school resources co-created by computer science academics and teachers. In: Proc. of the 27th ACM Conference on on Innovation and Technology in Computer Science Education Vol. 1. pp. 207--213. ACM (2022). \doi{10.1145/3502718.3524804}

\bibitem{Luu2023}
Luu, M., Ferland, M., Rao, V.N., Arora, A., Huynh, R., Reiber, F., Wong-Ma, J., Shindler, M.: What is an algorithms course? In: Proc. of the 54th SIGCSE TS V. 1. pp. 284--290. ACM (2023). \doi{10.1145/3545945.3569820}

\bibitem{Luxton-Reilly2023}
Luxton-Reilly, A., Tempero, E., Arachchilage, N., Chang, A., Denny, P., Fowler, A., Giacaman, N., Kontorovich, I., Lottridge, D., Manoharan, S., Sindhwani, S., Singh, P., Speidel, U., Stephen, S., Terragni, V., Whalley, J., Wuensche, B., Ye, X.: Automated assessment: Experiences from the trenches. In: Proc. of the 25th Australasian Computing Education Conference. pp. 1--10. ACM (2023). \doi{10.1145/3576123.3576124}

\bibitem{Macann2023}
Macann, V., Yadav, A.: Debugging beyond the code: Teachers' perceptions of debugging as a ct practice impacting interdisciplinary teaching and learning. In: Proc. of the ACM Conference on Global Computing Education Vol 1. pp. 119--125. ACM (2023). \doi{10.1145/3576882.3617919}

\bibitem{Mahon2024}
Mahon, J., Namee, B.M., Becker, B.A.: Guidelines for the evolving role of generative ai in introductory programming based on emerging practice. In: Proc. of the 2024 on Innovation and Technology in Computer Science Education V. 1. pp. 10--16. ACM (2024). \doi{10.1145/3649217.3653602}

\bibitem{Mahoney2023}
Mahoney, M.: Storyteller: Guiding students through code examples. In: Proc. of the 54th SIGCSE TS V. 1. pp. 1131--1135. ACM (2023). \doi{10.1145/3545945.3569843}

\bibitem{Malik2024}
Malik, A., Woodrow, J., Wang, C., Piech, C.: Teachnow: Enabling teachers to provide spontaneous, realtime 1:1 help in massive online courses. In: Proc. of the 2024 on Innovation and Technology in Computer Science Education V. 1. pp. 708--714. ACM (2024). \doi{10.1145/3649217.3653629}

\bibitem{Markel2021}
Markel, J.M., Guo, P.J.: Inside the mind of a cs undergraduate ta. In: Proc. of the 52nd SIGCSE TS. pp. 502--508. ACM (2021). \doi{10.1145/3408877.3432533}

\bibitem{Marshall2020}
Marshall, B., Geier, S.: Cross-disciplinary faculty development in data science principles for classroom integration. In: Proc. of the 51st SIGCSE TS. pp. 1207--1213. ACM (2020). \doi{10.1145/3328778.3366801}

\bibitem{Mason2024}
Mason, R., Simon, Becker, B.A., Crick, T., Davenport, J.H.: A global survey of introductory programming courses. In: Proc. of the 55th SIGCSE TS V. 1. pp. 799--805. ACM (2024). \doi{10.1145/3626252.3630761}

\bibitem{McGill2023}
McGill, M.M., Bell, A., Baskin, J., Reinking, A., Sweet, M.: Measuring teacher growth based on the csta k-12 standards for cs teachers. In: Proc. of the 54th SIGCSE TS V. 1. pp. 994--1000. ACM (2023). \doi{10.1145/3545945.3569796}

\bibitem{McGill2024}
McGill, M.M., Tise, J.C., Decker, A.: Piloting a diagnostic tool to measure ap cs principles teachers' knowledge against csta teacher standard 1. In: Proc. of the 55th SIGCSE TS V. 1. pp. 819--825. ACM (2024). \doi{10.1145/3626252.3630905}

\bibitem{Minnes2022}
Minnes, M.: Designing ta training for computer science graduate students: Remote and self-paced options for a supported introduction to reflective teaching. In: Proc. of the 53rd SIGCSE TS. pp. 752--758. ACM (2022). \doi{10.1145/3478431.3499342}

\bibitem{Munasinghe2021}
Munasinghe, B., Bell, T., Robins, A.: Teachers’ understanding of technical terms in a computational thinking curriculum. In: Proc. of the 23rd Australasian Computing Education Conference. pp. 106--114. ACM (2021). \doi{10.1145/3441636.3442311}

\bibitem{Musaeus2024}
Musaeus, L.H., Petersen, M.G., Klokmose, C.N.: Bringing teachers and researchers together through participatory design and cooperative prototyping in computing education. In: Proc. of the 55th SIGCSE TS V. 1. pp. 902--908. ACM (2024). \doi{10.1145/3626252.3630796}

\bibitem{Ni2021}
Ni, L., McKlin, T., Hao, H., Baskin, J., Bohrer, J., Tian, Y.: Understanding professional identity of computer science teachers: Design of the computer science teacher identity survey. In: Proc. of the 17th ACM Conference on International Computing Education Research. pp. 281--293. ACM (2021). \doi{10.1145/3446871.3469766}

\bibitem{Nitta2022}
Nitta, C., Kaloti, A., Wang, S.: Risc-v console. In: Proc. of the 27th ACM Conference on on Innovation and Technology in Computer Science Education Vol. 1. pp. 145--150. ACM (2022). \doi{10.1145/3502718.3524791}

\bibitem{Nugent2022}
Nugent, G., Chen, K., Soh, L.K., Choi, D., Trainin, G., Smith, W.: Developing k-8 computer science teachers' content knowledge, self-efficacy, and attitudes through evidence-based professional development. In: Proc. of the 27th ACM Conference on on Innovation and Technology in Computer Science Education Vol. 1. pp. 540--546. ACM (2022). \doi{10.1145/3502718.3524771}

\bibitem{Parthasarathy2024}
Parthasarathy, P.D., Joshi, S.: Teaching digital accessibility in computing education: Views of educators in india. In: Proc. of the 2024 ACM Conference on International Computing Education Research - Volume 1. pp. 222--232. ACM (2024). \doi{10.1145/3632620.3671122}

\bibitem{Patel2024}
Patel, P., Moz-Ruiz, D., Garcia, R., Chatterjee, A., Morreale, P., Burnett, M.: From workshops to classrooms: Faculty experiences with implementing inclusive design principles. In: Proc. of the 55th SIGCSE TS V. 1. pp. 1035--1041. ACM (2024). \doi{10.1145/3626252.3630861}

\bibitem{Perlman2021}
Perlman, R., Cohen, H., Hazzan, O.: The cs-orona initiative. In: Proc. of the 52nd SIGCSE TS. pp. 732--738. ACM (2021). \doi{10.1145/3408877.3432474}

\bibitem{Pieper2020}
Pieper, U., Vahrenhold, J.: Critical incidents in k-12 computer science classrooms - towards vignettes for computer science teacher training. In: Proc. of the 51st SIGCSE TS. pp. 978--984. ACM (2020). \doi{10.1145/3328778.3366926}

\bibitem{Repenning2021}
Repenning, A., Lamprou, A., Basawapatna, A.: Computing effect sizes of a science-first-then-didactics computational thinking module for preservice elementary school teachers. In: Proc. of the 52nd SIGCSE TS. pp. 274--280. ACM (2021). \doi{10.1145/3408877.3432446}

\bibitem{Riese2022}
Riese, E., Kann, V.: Training teaching assistants by offering an introductory course. In: Proc. of the 53rd SIGCSE TS. pp. 745--751. ACM (2022). \doi{10.1145/3478431.3499270}

\bibitem{Riese2021}
Riese, E., Lorås, M., Ukrop, M., Effenberger, T.: Challenges faced by teaching assistants in computer science education across europe. In: Proc. of the 26th ITiCSE V. 1. pp. 547--553. ACM (2021). \doi{10.1145/3430665.3456304}

\bibitem{Rosenbloom2023}
Rosenbloom, L.N.: A living framework for abolitionist teaching in computer science. In: Proc. of the ACM Conference on Global Computing Education Vol 1. pp. 133--139. ACM (2023). \doi{10.1145/3576882.3617923}

\bibitem{Rusak2021}
Rusak, G., Yan, L.: Unique exams. In: Proc. of the 52nd SIGCSE TS. pp. 1170--1176. ACM (2021). \doi{10.1145/3408877.3432556}

\bibitem{Samarasekara2023}
Samarasekara, C.K., Ott, C., Robins, A.: Future scenarios for high school digital technology in new zealand. In: Proc. of the 25th Australasian Computing Education Conference. pp. 21--30. ACM (2023). \doi{10.1145/3576123.3576126}

\bibitem{Saraiva2021}
Saraiva, J., Zong, Z., Pereira, R.: Bringing green software to computer science curriculum: Perspectives from researchers and educators. In: Proc. of the 26th ITiCSE V. 1. pp. 498--504. ACM (2021). \doi{10.1145/3430665.3456386}

\bibitem{Sauerwein2023}
Sauerwein, C., Antensteiner, T., Oppl, S., Groher, I., Meschtscherjakov, A., Zech, P., Breu, R.: Towards a success model for automated programming assessment systems used as a formative assessment tool. In: Proc. of the 2023 Conference on Innovation and Technology in Computer Science Education V. 1. pp. 271--277. ACM (2023). \doi{10.1145/3587102.3588848}

\bibitem{Sentance2021}
Sentance, S., Waite, J.: Teachers’ perspectives on talk in the programming classroom : Language as a mediator. In: Proc. of the 17th ACM Conference on International Computing Education Research. pp. 266--280. ACM (2021). \doi{10.1145/3446871.3469751}

\bibitem{Shanley2023}
Shanley, N., Pérez-Quiñones, M.A., Martin, F., Pugalee, D., Ahlgrim-Delzell, L., Hart, E.: K-12 teacher experiences from online professional development for teaching apcsa. In: Proc. of the 54th SIGCSE TS V. 1. pp. 1001--1006. ACM (2023). \doi{10.1145/3545945.3569827}

\bibitem{Sheard2024}
Sheard, J., Denny, P., Hellas, A., Leinonen, J., Malmi, L., Simon: Instructor perceptions of ai code generation tools - a multi-institutional interview study. In: Proc. of the 55th SIGCSE TS V. 1. pp. 1223--1229. ACM (2024). \doi{10.1145/3626252.3630880}

\bibitem{Shrestha2022}
Shrestha, R., Leinonen, J., Hellas, A., Ihantola, P., Edwards, J.: Codeprocess charts: Visualizing the process of writing code. In: Proc. of the 24th Australasian Computing Education Conference. pp. 46--55. ACM (2022). \doi{10.1145/3511861.3511867}

\bibitem{Simmonds2021}
Simmonds, J., Gutierrez, F.J., Meza, F., Torrent, C., Villalobos, J.: Changing teacher perceptions about computational thinking in grades 1-6, through a national training program. In: Proc. of the 52nd SIGCSE TS. pp. 260--266. ACM (2021). \doi{10.1145/3408877.3432542}

\bibitem{Singh2020}
Singh, S., Meyer, B., Wybrow, M.: Userflow: A tool for visualizing fine-grained contextual analytics in teaching documents. In: Proc. of the 2020 ITiCSE. pp. 384--390. ACM (2020). \doi{10.1145/3341525.3387410}

\bibitem{Skuratowicz2021}
Skuratowicz, E., Vanderberg, M., Hung, E.E., Krause, G., Bradley, D., Wilson, J.P.: I felt like we were actually going somewhere. In: Proc. of the 52nd SIGCSE TS. pp. 739--745. ACM (2021). \doi{10.1145/3408877.3432482}

\bibitem{Su2021}
Su, S., Zhang, E., Denny, P., Giacaman, N.: A game-based approach for teaching algorithms and data structures using visualizations. In: Proc. of the 52nd SIGCSE TS. pp. 1128--1134. ACM (2021). \doi{10.1145/3408877.3432520}

\bibitem{Thigpen2024}
Thigpen, L., McGill, M.M., Twarek, B., Bell, A.: Piloting a revised diagnostic tool for csta standards for cs teachers. In: Proc. of the 2024 on ACM Virtual Global Computing Education Conference V. 1. pp. 207--213. ACM (2024). \doi{10.1145/3649165.3690122}

\bibitem{Tsan2022}
Tsan, J., Coenraad, M., Crenshaw, Z., Palmer, J., Eatinger, D., Beck, K., Weintrop, D., Franklin, D.: Reimagining professional development for k-8 cs teachers. In: Proc. of the 53rd SIGCSE TS. pp. 530--536. ACM (2022). \doi{10.1145/3478431.3499361}

\bibitem{Tsan2022b}
Tsan, J., Weintrop, D., Franklin, D.: An analysis of middle grade teachers' debugging pedagogical content knowledge. In: Proc. of the 27th ACM Conference on on Innovation and Technology in Computer Science Education Vol. 1. pp. 533--539. ACM (2022). \doi{10.1145/3502718.3524770}

\bibitem{Tshukudu2021}
Tshukudu, E., Cutts, Q., Goletti, O., Swidan, A., Hermans, F.: Teachers’ views and experiences on teaching second and subsequent programming languages. In: Proc. of the 17th ACM Conference on International Computing Education Research. pp. 294--305. ACM (2021). \doi{10.1145/3446871.3469752}

\bibitem{Upadhyaya2020}
Upadhyaya, B., McGill, M.M., Decker, A.: A longitudinal analysis of k-12 computing education research in the united states. In: Proc. of the 51st SIGCSE TS. pp. 605--611. ACM (2020). \doi{10.1145/3328778.3366809}

\bibitem{Vahid2024}
Vahid, F.: Cs1 instructors: Flexibility in content approaches is justified, and can enable more cross-university cooperation. In: Proc. of the 55th SIGCSE TS V. 1. pp. 1368--1373. ACM (2024). \doi{10.1145/3626252.3630816}

\bibitem{Vahid2024b}
Vahid, F., Pang, A.: Experiences teaching a cs1 common course across 7 institutions. In: Proc. of the 55th SIGCSE TS V. 1. pp. 1354--1360. ACM (2024). \doi{10.1145/3626252.3630847}

\bibitem{Valstar2020}
Valstar, S., Sih, C., Krause-Levy, S., Porter, L., Griswold, W.G.: A quantitative study of faculty views on the goals of an undergraduate cs program and preparing students for industry. In: Proc. of the 2020 ACM Conference on International Computing Education Research. pp. 113--123. ACM (2020). \doi{10.1145/3372782.3406277}

\bibitem{VanDeGrift2023}
VanDeGrift, T.: Alumni as teachers and mentors for cs 1 students. In: Proc. of the 54th SIGCSE TS V. 1. pp. 1124--1130. ACM (2023). \doi{10.1145/3545945.3569721}

\bibitem{Venn-Wycherley2020}
Venn-Wycherley, M., Bennett, C., Kharrufa, A.: Design studios for k-12 computing education. In: Proc. of the 51st SIGCSE TS. pp. 1227--1233. ACM (2020). \doi{10.1145/3328778.3366923}

\bibitem{Vivian2020}
Vivian, R., Quille, K., McGill, M.M., Falkner, K., Sentance, S., Barksdale, S., Busuttil, L., Cole, E., Liebe, C., Maiorana, F.: An international pilot study of k-12 teachers' computer science self-esteem. In: Proc. of the 2020 ITiCSE. pp. 117--123. ACM (2020). \doi{10.1145/3341525.3387418}

\bibitem{Wang2021}
Wang, W., Zhang, C., Stahlbauer, A., Fraser, G., Price, T.: Snapcheck: Automated testing for snap <i>!</i> programs. In: Proc. of the 26th ITiCSE V. 1. pp. 227--233. ACM (2021). \doi{10.1145/3430665.3456367}

\bibitem{Weintrop2022}
Weintrop, D.: ischools as venues for expanding the k-12 computer science teacher pipeline. In: Proc. of the 53rd SIGCSE TS. pp. 397--403. ACM (2022). \doi{10.1145/3478431.3499302}

\bibitem{Wilde2024}
Wilde, J., Beltran, E., Zawatski, M.J., Fernandez, A.S., Prasad, P.V., Yuen, T.T.: Experiences in delivering online cs teacher professional development. In: Proc. of the 55th SIGCSE TS V. 1. pp. 1428--1434. ACM (2024). \doi{10.1145/3626252.3630845}

\bibitem{Williams2020}
Williams, H.E., Williams, S., Kendall, K.: Cs in schools: Developing a sustainable coding programme in australian schools. In: Proc. of the 2020 ITiCSE. pp. 321--327. ACM (2020). \doi{10.1145/3341525.3387422}

\bibitem{Yadav2022}
Yadav, A., Caeli, E.N., Ocak, C., Macann, V.: Teacher education and computational thinking. In: Proc. of the 27th ACM Conference on on Innovation and Technology in Computer Science Education Vol. 1. pp. 547--553. ACM (2022). \doi{10.1145/3502718.3524783}

\bibitem{Yadav2022Review}
Yadav, A., Connolly, C., Berges, M., Chytas, C., Franklin, C., Hijón-Neira, R., Macann, V., Margulieux, L., Ottenbreit-Leftwich, A., Warner, J.R.: A review of international models of computer science teacher education. In: Proc. of the 2022 Working Group Reports on Innovation and Technology in Computer Science Education. pp. 65--93. ACM (2022). \doi{10.1145/3571785.3574123}

\bibitem{Zaman2023}
Zaman, A., Cook, A., Phan, V., Windsor, A.: A practical strategy for training graduate cs teaching assistants to provide effective feedback. In: Proc. of the 2023 Conference on Innovation and Technology in Computer Science Education V. 1. pp. 285--291. ACM (2023). \doi{10.1145/3587102.3588776}

\end{thebibliography}

\end{document}